\documentclass[aps,prd,amsmath
,eqsecnum            
,preprintnumbers
,nofootinbib         
 ,showpacs          
 ,showkeys          
,twocolumn         
]{revtex4}
\usepackage[dvips]{graphicx}
\usepackage{amsmath}
\usepackage{amsfonts}
\usepackage{amssymb}
\usepackage{longtable}   

\allowdisplaybreaks[4]     

\newcommand{\df}{\ {\overset {\rm def} =}\ }
\newcommand{\dr}[2]{\frac {{\rm d} {#1}} {{\rm d} {#2}}}

\newcommand{\dril}[2]{{{\rm d} {#1}} / {{\rm d} {#2}}}

\begin{document}

\title{Accelerating expansion or inhomogeneity? Part 2: \\
Mimicking acceleration with the energy function in the Lema\^{\i}tre -- Tolman
model}

\author{Andrzej Krasi\'nski}
\affiliation{N. Copernicus Astronomical Centre, Polish Academy of Sciences, \\
Bartycka 18, 00 716 Warszawa, Poland} \email{akr@camk.edu.pl}

\date {}

\begin{abstract}
This is a continuation of the paper published in {\it Phys. Rev.} {\bf D89},
023520 (2014). It is investigated here how the luminosity distance -- redshift
relation $D_L(z)$ of the $\Lambda$CDM model is duplicated in the Lema\^{\i}tre
-- Tolman (L--T) model with $\Lambda = 0$, constant bang-time function $t_B$ and
the energy function $E(r)$ mimicking accelerated expansion on the observer's
past light cone ($r$ is a uniquely defined comoving radial coordinate).
Numerical experiments show that $E > 0$ necessarily. The functions $z(r)$ and
$E(r)$ are numerically calculated from the initial point at the observer's
position, then backward from the initial point at the apparent horizon (AH).
Reconciling the results of the two calculations allows one to determine the
values of $E/r^2$ at $r = 0$ and at the AH. The problems connected with
continuing the calculation through the AH are discussed in detail and solved.
Then $z(r)$ and $E(r)$ are continued beyond the AH, up to the numerical crash
that signals the contact of the light cone with the Big Bang. Similarly, the
light cone of the L--T model is calculated by proceeding from the two initial
points, and compared with the $\Lambda$CDM light cone. The model constructed
here contains shell crossings, but they can be removed by matching the L--T
region to a Friedmann background, without causing any conflict with the type Ia
supernovae observations. The mechanism of imitating the accelerated expansion by
the $E(r)$ function is explained in a descriptive way.
\end{abstract}

\maketitle

\section{Introduction}\label{genintro}

\setcounter{equation}{0}

It is shown here how the luminosity distance -- redshift relation $D_L(z)$ of
the $\Lambda$CDM model is duplicated in the Lema\^{\i}tre \cite{Lema1933} --
Tolman \cite{Tolm1934} (L--T) model with $\Lambda = 0$, constant bang-time
function $t_B$ and the energy function $E(r)$ mimicking accelerated expansion on
the observer's past light cone. In such an L--T model there is no accelerated
expansion -- the $D_L(z)$ function results from a suitable inhomogeneous
distribution of matter in space.

This paper is a continuation of Ref. \cite{Kras2014}, where the duplication of
$D_L(z)$ was achieved using an L--T model with $\Lambda = 0$, constant $E/r^2 =
- k$, and $t_B$ mimicking the accelerated expansion. The studies in Ref.
\cite{Kras2014} and here were motivated by the paper by Iguchi, Nakamura and
Nakao \cite{INNa2002}, and are its extensions. In Ref. \cite{INNa2002}, just the
numerical proof of existence of such L--T models was given, but their geometry
was not discussed. The main purpose of this paper, along with Ref.
\cite{Kras2014}, is a deeper understanding of geometrical relations between the
two types of L--T models and the $\Lambda$CDM model -- in particular, of the
relation between their light cones.

As in Ref. \cite{Kras2014}, emphasis is put on analytical calculations;
numerical computations are postponed as much as possible. Formulae for the
limits of several quantities at $z \to 0$ are found; to a lesser extent this is
also possible for the limits at the apparent horizon (AH). This allows one to
verify the precision of some numerical calculations by carrying them out from
the initial point at $z = 0$, from the initial point at the AH, and comparing
the results.

The motivation and historical background were explained in Ref. \cite{Kras2014}.
Section \ref{basform} provides the basic formulae for reference. Its subsections
are condensed versions of sections II, III and VIII -- X of Ref.
\cite{Kras2014}. In Sec. \ref{LTwithnonzeroE}, the set of differential equations
defining $z(r)$ and $E(r)$ for the L--T model is derived. In Secs.
\ref{limits0withE} and \ref{limitsAH}, the limits of various quantities at $z
\to 0$ and at the AH are calculated. In Secs. \ref{deterXandk} --
\ref{numcalzFE} the equations for $z(r)$ and $E(r)$ are reformulated so as to
minimise the numerical instabilities in the vicinity of $r = 0$. In Sec.
\ref{negativeE} it is shown that the equations cannot be solved with $E \leq 0$.

In Sec. \ref{positiveE}, the equations are numerically solved with $E > 0$ by
proceeding from the initial point at $z = 0$. In Sec. \ref{verifyback}, the
solutions are found again by proceeding backward from the initial point at the
AH, and the two solutions are compared. The conditions that the $z(r)$ and
$E(r)$ curves calculated from $r = 0$ hit the points $(r, z) = (r_{\rm AH},
z_{\rm AH})$ and $(r, E) = (r_{\rm AH}, E_{\rm AH})$ determine the value of
$E/r^2$ at $r = 0$ and a provisional value of $E$ at $r = r_{\rm AH}$ (the
subscript ``AH'' denotes the value at the apparent horizon). The condition that
$E(r)$ calculated from the initial point at $r = r_{\rm AH}$ hits $(r, E) = (0,
0)$ allows us to calculate a corrected value of $E$ at the AH.

In Sec. \ref{continuepastAH}, the $z(r)$ and $E(r)$ curves are extended by
proceeding forward from the initial point at the AH up to the numerical crash
that signals the contact of the light cone with the Big Bang (BB). It turns out
that $E(r)$ becomes decreasing at $r = r_{\rm sc} > r_{\rm AH}$, so there are
shell crossings at $r > r_{\rm sc}$. The region containing shell crossings can
be removed from the model by matching the L--T solution to a Friedmann
background across a hypersurface $r = r_B =$ constant $< r_{\rm sc}$. The
redshift corresponding to $r_{\rm sc}$ is $z_{\rm sc} = 6.938073260172738$, so
the matching surface can be farther from the observer than the type Ia
supernovae -- see Sec. \ref{continuepastAH} for more on this.

In Sec. \ref{lightcone}, the past light cone of the central observer in the L--T
model is calculated by proceeding from $r = 0$ and by proceeding backward from
$r = r_{\rm AH}$. Consistency between these calculations is satisfactory. Then,
the calculation is continued up to the BB. The L--T light cone is compared with
that of the $\Lambda$CDM model.

In Sec. \ref{conclu}, the imitation of the accelerated expansion by the $E(r)$
function is explained in a descriptive way, and the conclusions are presented.
One of them is that the value of $k = \lim_{r \to 0} (- 2E/r^2)$ is fixed by the
values of $r$ and $z$ at the AH, which, in turn, are fixed by the
observationally determined parameters of the $\Lambda$CDM model: the Hubble
constant ${\cal H}_0$ and the density and cosmological constant parameters
$\Omega_m$ and $\Omega_{\Lambda}$. Consequently, $k$ cannot be treated as a free
parameter to be adjusted to observations, as was done in some of the earlier
papers.

\section{Basic formulae}\label{basform}

\setcounter{equation}{0}

\subsection{An introduction to the L--T models}\label{LTintro}

This is a summary of basic facts about the L--T model. For extended expositions
see Refs. \cite{PlKr2006,Kras1997}. Its metric is:
\begin{equation}\label{2.1}
{\rm d} s^2 = {\rm d} t^2 - \frac {{R_{,r}}^2}{1 + 2E(r)}{\rm d} r^2 -
R^2(t,r)({\rm d}\vartheta^2 + \sin^2\vartheta \, {\rm d}\varphi^2),
\end{equation}
where $E(r)$ is an arbitrary function, and $R(t, r)$ is determined by the
integral of the Einstein equations:
\begin{equation}\label{2.2}
{R_{,t}}^2 = 2E(r) + 2M(r) / R - \tfrac 1 3 \Lambda R^2,
\end{equation}
$M(r)$ being another arbitrary function and $\Lambda$ being the cosmological
constant. Note that $E$ must obey
\begin{equation}\label{2.3}
2E + 1 \geq 0
\end{equation}
in order that the signature of (\ref{2.1}) is $(+ - - -)$.

In the case $\Lambda = 0$, the solutions of (\ref{2.2}) are:

(1) When $E(r) < 0$:
\begin{eqnarray}\label{2.4}
R(t,r) &=& - \frac M {2E} (1 - \cos \eta), \nonumber \\
\eta - \sin \eta &=& \frac {(-2E)^{3/2}} M \left[t - t_B(r)\right].
\end{eqnarray}

(2) When $E(r) = 0$:
\begin{equation}\label{2.5}
R(t,r) = \left\{\frac 9 2 M(r) \left[t - t_B(r)\right]^2\right\}^{1/3}.
\end{equation}

(3) When $E(r) > 0$:
\begin{eqnarray}\label{2.6}
R(t,r) &=& \frac M {2E} (\cosh \eta - 1), \nonumber \\
\sinh \eta - \eta &=& \frac {(2E)^{3/2}} M \left[t - t_B(r)\right].
\end{eqnarray}
The case $E(r) = 0$ can occur either in a 4-dimensional region or on a
3-dimensional boundary between $E > 0$ and $E < 0$ regions, at a single value of
$r$ -- but it will not occur in this paper.

The pressure is zero, so the matter (dust) particles move on geodesics. The mass
density is
\begin{equation}  \label{2.7}
\kappa \rho = \frac {2{M_{,r}}}{R^2R_{,r}}, \qquad \kappa \df \frac {8\pi G}
{c^2}.
\end{equation}

The coordinate $r$ in (\ref{2.1}) is determined up to arbitrary transformations
of the form $r = f(r')$. This freedom allows us to give one of the functions
$(M, E, t_B)$ a handpicked form (under suitable assumptions that guarantee
uniqueness of the transformation). We make $r$ unique by assuming $M,_r > 0$ and
choosing $r$ as follows:
\begin{equation}\label{2.8}
M = M_0 r^3,
\end{equation}
with $M_0 = 1$. This value of $M_0$ can be obtained by the transformation $r =
Cr'$, $C =$ constant. Choosing a value for $M_0$ is equivalent to choosing a
unit of mass \cite{Kras2014}.

A past radial null geodesic is given by the equation
\begin{equation}\label{2.9}
\dr t r = - \frac {R_{,r}} {\sqrt{1 + 2E(r)}},
\end{equation}
and its solution is denoted $t = t_{\rm ng}(r)$. The redshift $z(r)$ along
$t_{\rm ng}(r)$ is given by \cite{Bond1947,PlKr2006}
\begin{equation}\label{2.10}
\frac 1 {1 + z}\ \dr z r = \left[ \frac {R_{,tr}} {\sqrt{1 + 2E}} \right]_{\rm ng}.
\end{equation}
Given $t_{\rm ng}(r)$ and $z(r)$, the luminosity distance $D_L(z)$ of a light
source from the central observer is \cite{BKHC2010}
\begin{equation}\label{2.11}
D_L(z) = (1 + z)^2 \left.R\right|_{\rm ng}.
\end{equation}

\subsection{The Friedmann limit of the L--T model, the $\Lambda$CDM model}

The Friedmann limit of (\ref{2.1}) follows when $M/r^3 = M_0$, $2E / r^2 = - k$
and $t_B$ are constant, where $k$ is the Friedmann curvature index. Then
(\ref{2.4}) -- (\ref{2.6}) imply $R = r S(t)$, and the limiting metric is
\begin{equation}\label{2.12}
{\rm d} s^2 = {\rm d} t^2 - S^2(t) \left[\frac 1 {1 - kr^2} {\rm d} r^2 +
r^2 ({\rm d}\vartheta^2 + \sin^2\vartheta \, {\rm d}\varphi^2)\right].
\end{equation}
Equation (\ref{2.10}) is easily integrated to give
\begin{equation}\label{2.13}
1 + z = S(t_o)/S(t_e),
\end{equation}
where $t_o$ and $t_e$ are the instants of, respectively, the observation and
emission of the light ray.

The $\Lambda$CDM model is a solution of Einstein's equations for the metric
(\ref{2.12}) with dust source and $k = 0 > \Lambda$ \cite{Kras2014}:
\begin{equation}\label{2.14}
S(t) = \left(- \frac {6M_0} {\Lambda}\right)^{1/3} \sinh^{2/3} \left[\frac
{\sqrt {- 3 \Lambda}} 2 \left(t - t_{B\Lambda}\right)\right],
\end{equation}
where $t = t_{B\Lambda}$ is the instant of the BB. The $D_L(z)$ formula in this
model can be represented as follows:
\begin{equation}\label{2.15}
D_L(z) = \frac {1 + z} {H_0} \int_0^z \frac {{\rm d} z'} {\sqrt{\Omega_m (1 +
z')^3 + \Omega_{\Lambda}}},
\end{equation}
where $H_0$ is the Hubble parameter at $t_o$,
\begin{equation}\label{2.16}
H_0 = \left.S,_t/S\right|_{t = t_o}
\end{equation}
and the two dimensionless parameters
\begin{equation}\label{2.17}
\left(\Omega_m, \Omega_{\Lambda}\right) \df \frac 1 {3{H_0}^2}
\left.\left(\frac {8\pi G \rho_0} {c^2}, - \Lambda\right)\right|_{t = t_o}
\end{equation}
obey $\Omega_m + \Omega_{\Lambda} \equiv 1$; $\rho_0$ is the present mean mass
density in the Universe. Equation (\ref{2.15}) follows by combining (\ref{2.11})
with (\ref{2.9}) and (\ref{2.2}) in the $\Lambda$CDM limit, where $E = k = 0$.

The Hubble parameter $H_0$ in (\ref{2.16}) is related to the Hubble constant
${\cal H}_0 = 67.1$ km/(s $\times$ Mpc) \cite{Plan2013} by
\begin{equation}\label{2.18}
H_0 = {\cal H}_0 / c.
\end{equation}

\subsection{Regularity conditions}\label{regconds}

Two kinds of singularity may occur in the L--T models apart from the BB: shell
crossings \cite{HeLa1985}, \cite{PlKr2006} and a permanent central singularity
\cite{PlKr2006}.

With the assumptions $M,_r > 0$ and $t_{B,r} = 0$ adopted here, the necessary
and sufficient conditions for the absence of shell crossings are \cite{HeLa1985}
\begin{eqnarray}
\frac {M,_r} M &>& \frac {3 E,_r} {2E}, \qquad {\rm when\ } E < 0, \label{2.19}
\\
E,_r &>& 0 \qquad {\rm when\ } E > 0. \label{2.20}
\end{eqnarray}

To avoid a permanent central singularity, the function $E$ must have the form
\cite{PlKr2006}
\begin{equation}\label{2.21}
2E = r^2 \left(- k + {\cal F}(r)\right),
\end{equation}
where $k =$ constant (possibly 0) and
\begin{equation}\label{2.22}
\lim_{r \to 0} {\cal F} = 0.
\end{equation}

\subsection{Apparent horizons in the L--T and Friedmann models}\label{AHs}

The AH of the central observer is a locus where $R$, calculated along a
past-directed null geodesic given by (\ref{2.9}), changes from increasing to
decreasing, i.e., where
\begin{equation}\label{2.23}
\dr {} r R(t_{\rm ng}(r), r) = 0.
\end{equation}
This locus is given by \cite{KrHe2004}
\begin{equation}\label{2.24}
2M / R - 1 - \tfrac 1 3 \Lambda R^2 = 0.
\end{equation}
Equation (\ref{2.24}) has a unique $R > 0$ solution for every value of $\Lambda$
(see Appendix A of Ref. \cite{Kras2014}). Thus, the AH exists independently of
the value of $\Lambda$. The same applies to the Friedmann models
\cite{Elli1971}.

{}From now on, $\Lambda = 0$ will be assumed for the L--T model, so the AH will
be at
\begin{equation}\label{2.25}
R = 2M = 2M_0 r^3.
\end{equation}

\subsection{Duplicating the luminosity distance -- redshift relation using the
L--T model with $\Lambda = 0$}\label{duplicate}

To duplicate (\ref{2.15}) using the $\Lambda = 0$ L--T model means, in view of
(\ref{2.11}), to require that
\begin{equation}\label{2.26}
R(t_{\rm ng}(r), r) = \frac 1 {H_0 (1 + z)} \int_0^z \frac {{\rm d} z'}
{\sqrt{\Omega_m (1 + z')^3 + \Omega_{\Lambda}}}
\end{equation}
holds along the past light cone of the central observer, where $H_0, \Omega_m$
and $\Omega_{\Lambda}$ have the values determined by current observations
\cite{Plan2013}, $t_{\rm ng}(r)$ is the function determined by (\ref{2.9}) and
$z(r)$ is determined by (\ref{2.10}). Let
\begin{equation}\label{2.27}
{\cal D}(z) \df \int_0^z \frac {{\rm d} z'} {\sqrt{\Omega_m (1 + z')^3 +
\Omega_{\Lambda}}}.
\end{equation}
Note that ${\cal D}(0) = 0$, ${\cal D}(z) > 0$ at all $z > 0$ and ${\cal D},_z >
0$ at all $z \geq 0$, but $\lim_{z \to \infty}{\cal D}(z)$ is finite.

Light emitted at the BB of an L--T model is, in general, infinitely blueshifted,
i.e. $z_{\rm BB} = -1$, except when $t_{B,r} = 0$ at the emission point
\cite{Szek1980}, \cite{HeLa1984}, \cite{PlKr2006}. Since we consider here the
L--T model with constant $t_B$, all light emitted at the BB will be infinitely
redshifted, just as in the Robertson -- Walker (RW) models. This is seen from
(\ref{2.26}): since $0 < {\cal D} < \infty$ for all $z > 0$ and $R = 0$ at the
BB, $z \to \infty$ must hold at the BB.

\subsection{Locating the apparent horizon}\label{locateAH}

Differentiating (\ref{2.26}) by $r$ and using (\ref{2.23}) one obtains
\begin{equation}\label{2.28}
\left.A_1\right|_{\rm AH} = 0,
\end{equation}
where
\begin{equation}\label{2.29}
A_1 \df {\cal D} - \frac {1 + z} {\sqrt{\Omega_m (1 + z)^3 + \Omega_{\Lambda}}}.
\end{equation}

Equation (\ref{2.25}) may be written, using (\ref{2.26}), (\ref{2.27}) and
(\ref{2.8}), also as
\begin{equation}\label{2.30}
r_{\rm AH} = \left[\frac {\cal D} {2M_0 H_0 (1 + z)}\right]^{1/3}_{\rm AH}.
\end{equation}

Note that (\ref{2.28}) does not refer to the parameters of the L--T model. So,
with $\Omega_m$ and $\Omega_{\Lambda}$ given, it can be numerically solved for
$z_{\rm AH}$ already at this stage, and the corresponding ${\cal D}_{\rm AH}$
and $r_{\rm AH}$ can be calculated from (\ref{2.27}) and (\ref{2.30}). The
solutions are the same as in Ref. \cite{Kras2014}:\footnote{The numbers
calculated for this paper by Fortran 90 are all at double precision -- to
minimise misalignments in the graphs.}

\begin{eqnarray}
z_{\rm AH} &=& 1.582430687623614, \label{2.31} \\
{\cal D}_{\rm AH} &=& 1.037876401742206, \label{2.32} \\
r_{\rm AH} &=& 0.3105427968086945. \label{2.33}
\end{eqnarray}

\subsection{The numerical units}\label{numerunits}

The following values are assumed here:
\begin{equation}\label{2.34}
(\Omega_m, \Omega_{\Lambda}, H_0, M_0) = (0.32, 0.68, 6.71, 1)
\end{equation}
the first two after Ref. \cite{Plan2013}. The $H_0$ is $1/10$ of the
observationally determined value of the Hubble constant \cite{Plan2013}
\begin{equation}\label{2.35}
{\cal H}_0 = c H_0 = 67.1\ {\rm km/(s} \times {\rm Mpc}).
\end{equation}
It follows that $H_0$ is measured in 1/Mpc. Consequently, choosing a value for
$H_0$ amounts to defining a numerical length unit; call it NLU. With
(\ref{2.34}), and assuming $c \approx 3 \times 10^5$ km/s, we have
\begin{equation}\label{2.36}
1\ {\rm NLU} = 3 \times 10^4\ {\rm Mpc}.
\end{equation}

Our time coordinate is $t = c \tau$, where $\tau$ is measured in time units, so
$t$ is measured in length units. So it is natural to take the NLU defined in
(\ref{2.36}) also as the numerical time unit (NTU). Taking the following
approximate values for the conversion factors \cite{unitconver}:
\begin{eqnarray}\label{2.37}
1\ {\rm pc} &=& 3.086 \times 10^{13}\ {\rm km}, \nonumber \\
1\ {\rm y} &=& 3.156 \times 10^7\ {\rm s},
\end{eqnarray}
the following relations result from (\ref{2.36}):
\begin{equation}\label{2.38}
1\ {\rm NTU} = 1\ {\rm NLU} = 9.26 \times 10^{23}\ {\rm km} = 9.8 \times
10^{10}\ {\rm y}.
\end{equation}
For the observationally determined age of the Universe \cite{Plan2013} we have
\begin{equation}\label{2.39}
T = 13.819 \times 10^9\ {\rm y} = 0.141\ {\rm NTU}.
\end{equation}

The mass associated to $M_0 = 1$ NLU in (\ref{2.34}) is $m_0 \approx 10^{54}$
kg, but it will appear only via $M_0$.

\section{The L--T model with $t_B =$ constant that duplicates the $D_L(z)$ of
(\ref{2.15})}\label{LTwithnonzeroE}

\setcounter{equation}{0}

The functional shape of $t_B$ might be determined by tying it to an additional
observable quantity, as was done in Ref. \cite{CBKr2010}. However, then the
equations defining $t_B$ and $E$ are coupled, and numerical handling becomes
instantly necessary. To keep things transparent, we follow the approach of Ref.
\cite{INNa2002} and consider separately the two complementary cases when $E(r)$
and $t_B(r)$ have their Friedmann forms, $-2E/r^2 = k =$ constant and $t_B =$
constant, respectively. The first case was investigated in Ref. \cite{Kras2014}.
Here, we consider the second case,
\begin{equation}\label{3.1}
t_B = {\rm constant}.
\end{equation}
The $M$ is chosen as in (\ref{2.8}). Using (\ref{3.1}), we have \cite{KrHe2004}
\begin{eqnarray}\label{3.2}
R,_r &=& \left(\frac {M,_r} M - \frac {E,_r} E\right)R \nonumber \\
&+& \left(\frac 3 2 \frac {E,_r} E - \frac {M,_r} M\right) \left(t -
t_B\right) R,_t.
\end{eqnarray}

The cases $E > 0$ and $E < 0$ have to be considered separately. Since we assumed
constant $t_B$, the case (\ref{2.5}) will not occur with $E \equiv 0$ because
this would be the $k = 0$ Friedmann model. The equality $E = 0$ might, in
principle, occur at isolated values of $r$ that define boundaries between the $E
> 0$ and $E < 0$ regions, but $E \leq 0$ will not occur in this paper -- see Sec.
\ref{negativeE}.

\subsection{$E > 0$}

We write (\ref{2.6}) in the form
\begin{equation}\label{3.3}
t - t_B = \frac M {(2E)^{3/2}} (\sinh \eta - \eta),
\end{equation}
and take it along a null geodesic, i.e. assume that the $t$ above is the $t(r)$
obeying (\ref{2.9}). There is a subtle point here: (\ref{3.3}) will be
differentiated along the null geodesic, so $\eta$, and $R$ defined by $\eta$ via
(\ref{2.6}), will be taken on the geodesic before they are differentiated. In
particular, $R$ will be replaced by (\ref{2.26}) before differentiation.
However, the $R,_r$ on the right-hand side of (\ref{2.9}) is calculated before
being taken along the null geodesic, so it will be replaced by (\ref{3.2}), and
(\ref{2.26}) will be used only after that.

The following formulae, derived from (\ref{2.6}), will be helpful:
\begin{eqnarray}
\sinh \eta &=& \sqrt{\left(\frac {2ER} M + 1\right)^2 - 1} \nonumber  \\
&\equiv& \frac {\sqrt{2E} R} M\ \sqrt{2E + \frac {2M} R} \equiv \frac {\sqrt{2E}
R} M\ R,_t, \label{3.4} \\
\dr {\eta_{\rm ng}} r &=& \frac 1 {\sinh \eta_{\rm ng}}\ \left[\frac {2E} M\
\frac {\cal D} {H_0 (1 + z)}\right],_r. \label{3.5}
\end{eqnarray}
We also introduce the following symbols, using (\ref{2.21}):
\begin{eqnarray} {\cal U} &\df& \frac {2ER_{\rm ng}} M + 1 \equiv \frac {{\cal
D} (- k + {\cal
F})} {M_0 H_0 r (1 + z)} + 1 \equiv \cosh \eta_{\rm ng}, \nonumber \\
&& \label{3.6}
\end{eqnarray}
so that
\begin{equation}\label{3.7} \eta_{\rm ng} = \ln \left({\cal U} + \sqrt{{\cal
U}^2 - 1}\right),
\end{equation}
and further
\begin{eqnarray}
B_1 &\df& \sqrt{2E + \frac {2M} {R_{\rm ng}}} \equiv r \sqrt{- k + {\cal F} +
\frac {2M_0 H_0 r (1 + z)} {\cal D}} \nonumber \\
&\equiv& \left.R,_t\right|_{\rm ng} \equiv \sqrt{\frac {M({\cal U} + 1)} {R_{\rm
ng}}}, \label{3.8} \\
B_2 &\df& 1 - \frac {B_1} {\sqrt{1 + 2E}} \label{3.9} \\
&\equiv& 1 - \frac {\sqrt{2E + 2M_0 H_0 r^3 (1 + z)/{\cal D}}} {\sqrt{1 + 2E}}
\label{3.10} \\
&\equiv& \frac {1 - 2 M_0 H_0 r^3 (1 + z)/{\cal D}} {1 + 2E + B_1 \sqrt{1 +
2E}},
\label{3.11} \\
B_3 &\df& B_1 \frac {3M_0 H_0 (1 + z)} {2 (- k + {\cal F})^{3/2}} \nonumber \\
&\times& \left[\sqrt{{\cal U}^2 - 1} - \ln \left({\cal U} + \sqrt{{\cal U}^2 -
1}\right)\right]. \label{3.12}
\end{eqnarray}
The three forms of $B_2$ are equivalent, but each of them is useful in a
different situation. For example, (\ref{3.11}) gives the best precision close to
the AH, where $B_2 = 0$ -- see the remark under (\ref{7.8}).

Now we differentiate (\ref{3.3}) along a radial null geodesic and use
(\ref{2.8}), (\ref{3.4}) -- (\ref{3.7}), (\ref{2.26}) and (\ref{2.27}),
obtaining
\begin{eqnarray}\label{3.13}
\left.\dr t r\right|_{\rm ng} &=& \frac 1 {H_0 (1 + z) B_1}\
\left[B_3\left(\frac 2 r - \frac {E,_r} E\right)\right. \nonumber \\
&+& \left.{\cal D} \left(\frac {E,_r} E - \frac 3 r\right) - \frac {A_1 z,_r}
{1 + z}\right].
\end{eqnarray}
On the other hand, from (\ref{2.9}), using (\ref{3.2}), (\ref{2.6}),
(\ref{2.8}), (\ref{3.4}), (\ref{3.7}), (\ref{2.26}) -- (\ref{2.27}), (\ref{2.2})
and (\ref{3.8}), we have
\begin{eqnarray}\label{3.14}
\left.\dr t r\right|_{\rm ng} &=& \frac 1 {H_0 (1 + z) \sqrt{1 + 2E}}\
\left[B_3\left(\frac 2 r - \frac {E,_r} E\right)\right. \nonumber \\
&+& \left.{\cal D} \left(\frac {E,_r} E - \frac 3 r\right)\right].
\end{eqnarray}
Equating (\ref{3.13}) to (\ref{3.14}) and using (\ref{3.9}) we obtain
\begin{equation}\label{3.15}
B_2 \left(\frac {{\cal D} - B_3} E\ \dr E r + \frac {2B_3 - 3 {\cal D}} r\right)
= \frac {A_1} {1 + z}\ \dr z r.
\end{equation}

Now, from (\ref{2.10}), using (\ref{3.2}), (\ref{2.2}), (\ref{3.8}) and
(\ref{3.12}):
\begin{eqnarray}\label{3.16}
\frac 1 {1 + z}\ \dr z r &=& \frac 1 {E \sqrt{1 + 2E}}\ \left[\frac {B_1} 2 -
\frac {M_0 H_0 r^3 (1 + z)B_3} {{\cal D}^2 B_1}\right] \dr E r \nonumber \\
&+& \frac {2M_0 H_0 r^2 (1 + z) B_3} {\sqrt{1 + 2E} {\cal D}^2 B_1}.
\end{eqnarray}
Solving (\ref{3.15}) and (\ref{3.16}) for $\dril z r$ and $\dril E r$ we obtain
\begin{eqnarray}
\frac 1 E\ \dr E r &=& \frac {B_5} {B_4}, \label{3.17} \\
\frac 1 {1 + z}\ \dr z r &=& \frac {3 {B_1}^2 \left(B_3 - {\cal D}\right) - 2E
B_3} {2 r \sqrt{1 + 2E} B_1 B_4}, \label{3.18}
\end{eqnarray}
where
\begin{eqnarray}
B_4 &\df& \frac {A_1} {\sqrt{1 + 2E} B_2}\ \left[\frac {B_1} 2 - \frac {M_0 H_0
r^3 (1 + z) B_3} {{\cal D}^2 B_1}\right] \nonumber \\
&+& B_3 - {\cal D}, \label{3.19} \\
B_5 &\df& \frac {2 B_3 - 3 {\cal D}} r - \frac {A_1} {B_2}\ \frac {2 M_0 H_0 r^2
(1 + z) B_3} {\sqrt{1 + 2E} {\cal D}^2 B_1}.\ \ \ \ \ \ \ \  \label{3.20}
\end{eqnarray}
Note that at the AH we have $A_1 = B_2 = 0$, so $\dril E r$ and $\dril z r$
involve expressions that become 0/0 there.

Since $E(0) = 0$ and $\dril E r(0) = 0$ (by (\ref{2.21}) -- (\ref{2.22})), eq.
(\ref{3.17}) cannot be solved numerically as given; $E$ has to be replaced by
${\cal F}$ using (\ref{2.21}). The result is
\begin{equation}\label{3.21}
\frac 1 {- k + {\cal F}}\ \dr {\cal F} r = \frac {B_5} {B_4} - \frac 2 r \equiv
\frac 1 {r B_4}\ \left(- {\cal D} - \frac {A_1 B_1} {B_2 \sqrt{1 + 2E}}\right).
\end{equation}

\subsection{$E < 0$}

Going through the same sequence of operations as for $E > 0$, we now use
\begin{equation}\label{3.22}
t - t_B = \frac M {(- 2E)^{3/2}} (\eta - \sin \eta)
\end{equation}
instead of (\ref{3.3}) and
\begin{eqnarray}
\sin \eta &=& \sqrt{1 - \left(\frac {2ER} M + 1\right)^2} \label{3.23} \\
&\equiv& \frac {\sqrt{-2E} R} M\ \sqrt{2E + \frac {2M} R} \equiv \frac
{\sqrt{-2E} R} M\ R,_t, \nonumber \\
\dr {\eta_{\rm ng}} r &=& \frac 1 {\sin \eta_{\rm ng}}\ \left[\frac {- 2E} M\
\frac {\cal D} {H_0 (1 + z)}\right],_r \label{3.24}
\end{eqnarray}
instead of (\ref{3.4}) -- (\ref{3.5}). The final result is similar to
(\ref{3.15}), except that ${\cal U}$ defined as in (\ref{3.6}) now obeys
\begin{equation}\label{3.25}
{\cal U} \equiv \cos \eta_{\rm ng},
\end{equation}
and instead of $B_3$, the following expression appears:
\begin{equation}\label{3.26}
{\widetilde B}_3 \df B_1 \frac {3M_0 H_0 (1 + z)} {2 (k - {\cal F})^{3/2}}
\left(\arccos {\cal U} - \sqrt{1 - {\cal U}^2}\right),
\end{equation}
where ${\cal U} \in [0, \pi]$ (the Universe is in the expansion phase). The
equations corresponding to (\ref{3.15}) and (\ref{3.16}) are now of the same
form, except that $B_3$ is replaced by ${\widetilde B}_3$. Consequently,
(\ref{3.17}) and (\ref{3.18}) result again, but with $B_3$ replaced by
${\widetilde B}_3$ also within $B_4$ and $B_5$.

\section{The limits of (\ref{3.17}) and (\ref{3.18}) at $r \to 0$}
\label{limits0withE}

\setcounter{equation}{0}

\subsection{$E > 0$}

We note that $\lim_{r \to 0} z = 0$ for physical reasons. Knowing this, we find
from (\ref{2.27}), using $\Omega_m + \Omega_{\Lambda} \equiv 1$,
\begin{equation}\label{4.1}
\lim_{r \to 0} \frac {\cal D} r = \lim_{r \to 0} \dr z r \df X.
\end{equation}
Anticipating that $X \neq 0$, so that $\lim_{r \to 0} \left(r^3 / {\cal
D}\right) = 0$, we then find from (\ref{2.29}), (\ref{3.6}) -- (\ref{3.12}),
(\ref{3.17}) -- (\ref{3.20}) and (\ref{2.21}) -- (\ref{2.22})
\begin{eqnarray}
\lim_{r \to 0} {\cal U} &=& 1 - \frac {k X} {M_0 H_0} \df {\cal U}_0,
\label{4.2} \\
- \lim_{r \to 0} A_1 &=& \lim_{r \to 0} B_2 = 1, \label{4.3} \\
\lim_{r \to 0} B_1 &=& \lim_{r \to 0} B_3 = \lim_{r \to 0} B_4 = 0, \label{4.4}
\\
\lim_{r \to 0} \frac {B_3} {B_1} &\df& \left(\frac {B_3} {B_1}\right)_0 =
\nonumber \\
&& \hspace{-2cm} \frac {3 M_0 H_0} {2 (- k)^{3/2}} \left[\sqrt{{{\cal U}_0}^2 -
1} - \ln \left({\cal U}_0 + \sqrt{{{\cal U}_0}^2 - 1}\right)\right], \label{4.5}
\\
\lim_{r \to 0} \frac {B_1} r &=& \sqrt{- k + \frac {2 M_0 H_0} X}, \label{4.6}
\\
\lim_{r \to 0} \frac {B_3} r &=& \sqrt{- k + \frac {2 M_0 H_0} X} \left(\frac
{B_3} {B_1}\right)_0, \label{4.7} \\
\lim_{r \to 0} B_5 &=& 2 \left(\sqrt{- k + \frac {2 M_0 H_0} X} + \frac {M_0
H_0} {X^2}\right) \left(\frac {B_3} {B_1}\right)_0 \nonumber \\
&-& 3 X, \label{4.8} \\
\lim_{r \to 0} \frac {B_4} r &=& \left(\sqrt{- k + \frac {2 M_0 H_0} X} + \frac
{M_0 H_0} {X^2}\right) \left(\frac {B_3} {B_1}\right)_0 \nonumber \\
&-& X - \frac 1 2 \sqrt{- k + \frac {2 M_0 H_0} X}. \label{4.9}
\end{eqnarray}

Using the above, the limit of (\ref{3.16}) at $r \to 0$ yields
\begin{equation}\label{4.10}
\lim_{r \to 0} \dr z r \equiv X = \sqrt{- k + \frac {2 M_0 H_0} X},
\end{equation}
see Appendix \ref{solve410} for a proof. This is equivalent to
\begin{equation}\label{4.11}
X^3 + kX - 2 M_0 H_0 = 0,
\end{equation}
the same equation as in Ref. \cite{Kras2014}. It is shown in Appendix C of Ref.
\cite{Kras2014} that (\ref{4.11}) has a unique solution for $X > 0$.

Taking the limit of (\ref{3.21}) at $r \to 0$ we obtain
\begin{eqnarray}\label{4.12}
&& \lim_{r \to 0} \dr {\cal F} r = - \frac k {\lim_{r \to 0} \left(B_4/r\right)}
\nonumber \\
&& \times \lim_{r \to 0} \left[\frac 1 {r^2}\ \left(- {\cal D} - \frac {A_1 B_1}
{B_2 \sqrt{1 + 2E}}\right)\right].
\end{eqnarray}
Since, from (\ref{4.3}) and (\ref{2.21}), $\lim_{r \to 0} \left(B_2 \sqrt{1 +
2E}\right) = 1$, eq. (\ref{4.12}), using (\ref{4.1}), (\ref{3.10}), (\ref{4.6})
and (\ref{4.10}), can be written as
\begin{eqnarray}\label{4.13}
&&\lim_{r \to 0} \dr {\cal F} r = \frac {kX} {\lim_{r \to 0}
\left(B_4/r\right)}\ \left\{- X\right. \nonumber \\
&&\ \ \ \  + \left.\lim_{r
\to 0} \left[\frac 1 r \left(\sqrt{1 + 2E} + A_1 B_1 / {\cal
D}\right)\right]\right\}.
\end{eqnarray}
Calculating this limit is tricky, so the derivation is presented in Appendix
\ref{der418}. The result is
\begin{equation}\label{4.14}
\lim_{r \to 0} \dr {\cal F} r = k \frac {\left(\tfrac 3 2\ \Omega_m - 1\right)
X^2 - M_0 H_0 / X} {2 \left(X + M_0 H_0 / X^2\right) \left(B_3/B_1\right)_0 -
3X}.
\end{equation}

\subsection{$E < 0$}

Equation (\ref{4.5}) in the case $E < 0$ is replaced by
\begin{eqnarray}
\lim_{r \to 0} \frac {\widetilde{B}_3} {B_1} &\df& \left(\frac
{\widetilde{B}_3} {B_1}\right)_0 = \nonumber \\
&& \hspace{-2cm} \frac {3 M_0 H_0} {2 k^{3/2}} \left(\arccos {\cal U}_0 -
\sqrt{1 - {{\cal U}_0}^2}\right). \label{4.15}
\end{eqnarray}
In Eqs. (\ref{4.4}), (\ref{4.7}) -- (\ref{4.9}) and (\ref{4.14}), $B_3$ must be
replaced by $\widetilde{B}_3$; the other equations in the set (\ref{4.3}) --
(\ref{4.14}) apply unchanged to the case $E < 0$.

\section{The limits of (\ref{3.17}) and (\ref{3.18}) at $r \to r_{\rm AH}$}
\label{limitsAH}

\setcounter{equation}{0}

Equations (\ref{2.27}), (\ref{2.28}) and (\ref{2.30}) provide explicit values of
$r$, $z$ and ${\cal D}$ at the AH, but it is not possible to calculate an
explicit expression for $E$ at $r = r_{\rm AH}$, and the value of $E(r_{\rm
AH})$ emerges only when (\ref{3.21}) is actually solved. Since (\ref{3.21}) and
(\ref{3.18}) depend on $E$, the expressions for $\dril z r$ and $\dril E r$ at
the AH cannot be calculated in advance, either.

As already mentioned below (\ref{3.18}), we have
\begin{equation}\label{5.1}
\left.A_1\right|_{\rm AH} = \left.B_2\right|_{\rm AH} = 0,
\end{equation}
so the only term in (\ref{3.17}), (\ref{3.18}) and (\ref{3.21}) that behaves
like 0/0 at the AH is $A_1/B_2$, and we obtain, using (\ref{2.29}) and
(\ref{2.30}),
\begin{equation}\label{5.2}
\lim_{r \to r_{\rm AH}} \frac {A_1} {B_2} = - \Omega_m \lim_{r \to r_{\rm AH}}
\left[r (1 + 2E) {\cal D}^3 \dr z r\right].
\end{equation}

Using (\ref{5.2}) in (\ref{3.18}), and taking into account that
$\left[B_1\right]_{\rm AH} = \left.\sqrt{1 + 2E}\right|_{\rm AH}$, one obtains
\begin{equation}\label{5.3}
\alpha \left(\dr z r\right)^2 + \beta \dr z r + \gamma = 0,
\end{equation}
where
\begin{eqnarray}
\alpha &=& \left\{\Omega_m r^2 (1 + 2E) {\cal D}^3 \left[1 + 2E - B_3 /
{\cal D}\right]\right\}_{\rm AH},\ \ \ \ \ \  \label{5.4} \\
\beta &=& \left\{- 2 r (1 + 2E) \left(B_3 - {\cal D}\right)\right\}_{\rm AH},
\label{5.5} \\
\gamma &=& \left\{(1 + z) \left[3 (1 + 2E) \left(B_3 - {\cal D}\right) - 2E
B_3\right]\right\}_{\rm AH}. \label{5.6}
\end{eqnarray}
Equation (\ref{5.3}) can be solved once the numerical value of $E(r_{\rm AH})$
is known. It will be calculated in Sec. \ref{verifyback}. With that value,
$\beta^2 - 4 \alpha \gamma > 0$, so (\ref{5.3}) has two real solutions. One of
them is negative, the other one is given by (\ref{10.2}).

With $E(r_{\rm AH})$ known, one more quantity can be calculated. From
(\ref{2.25}) and (\ref{2.6}), we have when $E > 0$
\begin{eqnarray}
&& \left.\cosh \eta\right|_{\rm AH} = 1 + 4 E(r_{\rm AH}) \df Y, \label{5.7} \\
&& \left(t - t_B\right)_{\rm AH} = \left[\frac {M_0 r^3}
{(2E)^{3/2}}\right]_{\rm AH} \nonumber \\
&&\ \ \ \ \ \times \left[\sqrt{Y^2 - 1} - \ln \left(Y + \sqrt{Y^2 -
1}\right)\right]. \label{5.8}
\end{eqnarray}

For $E < 0$ we have, from (\ref{2.25}) and (\ref{2.4}),
\begin{eqnarray}
&& \left.\cos \eta\right|_{\rm AH} = 1 + 4 E(r_{\rm AH}) \equiv Y, \label{5.9}
\\
&& \left(t - t_B\right)_{\rm AH} = \left[\frac {M_0 r^3} {(-
2E)^{3/2}}\right]_{\rm AH} \left(\arccos Y - \sqrt{1 - Y^2}\right), \nonumber
\\ \label{5.10}
\end{eqnarray}
where $0 \leq \eta \leq \pi$ (the Universe is in the expanding phase).

The past null geodesic of the central observer must pass through the point $(t,
r) = (t_{\rm AH}, r_{\rm AH})$, where $r_{\rm AH}$ is given by (\ref{2.33}) and
$t_{\rm AH}$ is given by (\ref{5.8}) or (\ref{5.10}). The numerical value of
$t_{\rm AH}$ can be calculated once the value of $E(r_{\rm AH})$ is known; it is
given by (\ref{10.3}).

\section{Determining $X$ and $k$}\label{deterXandk}

\setcounter{equation}{0}

The values of $k$ and $X$ are connected by (\ref{4.11}) and an equation derived
from (\ref{2.4}) (for $k > 0$) or (\ref{2.6}) (for $k < 0$), see below. Writing
(\ref{2.6}) in the form
\begin{equation}\label{6.1}
t - t_B = \frac M {(2E)^{3/2}} \left[\sqrt{{\cal U}^2 - 1} - \ln \left({\cal U}
+ \sqrt{{\cal U}^2 - 1}\right)\right],
\end{equation}
where ${\cal U}$ is given by (\ref{3.6}), we use (\ref{2.8}) and (\ref{2.21})
and take the limit of this at $r \to 0$. The result is
\begin{eqnarray}\label{6.2}
&& T_- \df \lim_{r \to 0} \left(t - t_B\right) = \nonumber \\
&& \frac {M_0} {(- k)^{3/2}}\ \left[\sqrt{{{\cal U}_0}^2 - 1} - \ln \left({\cal
U}_0 + \sqrt{{{\cal U}_0}^2 - 1}\right)\right],\ \ \ \ \ \
\end{eqnarray}
with ${\cal U}_0$ given by (\ref{4.2}) (the subscript ``minus'' refers to $k <
0$, which is equivalent to $X^3 > 2 M_0 H_0$). We have $\dril {T_-} X > 0$ at
all $X > (2M_0 H_0)^{1/3}$; see Appendix \ref{limproofs}. If $t$ is the present
instant, then $T_-$ is the age of the Universe in this model.

For $k > 0$ (i.e. $X^3 < 2 M_0 H_0$ and $E < 0$ in a neighbourhood of $r = 0$),
Eqs. (\ref{6.1}) and (\ref{6.2}) are replaced by
\begin{eqnarray}
&& t - t_B = \frac M {(-2E)^{3/2}} \left(\arccos {\cal U} - \sqrt{1-
{\cal U}^2}\right),\label{6.3} \\
&&T_+ \df \lim_{r \to 0} \left(t - t_B\right) = \frac {M_0} {k^{3/2}}\
\left(\arccos {\cal U}_0 - \sqrt{1 - {{\cal U}_0}^2}\right).\nonumber \\
&&\label{6.4}
\end{eqnarray}
Appendix \ref{limproofs} contains the proof that $\dril {T_+} X > 0$ for $0 <
X^3 < 2 M_0 H_0$ (i.e. $0 < k < \infty$).

It is tempting to assume $T_- = T$ or $T_+ = T$, where $T$ is given by
(\ref{2.39}), and then solve the set \{(\ref{6.2}), (\ref{4.11})\} or,
respectively, \{(\ref{6.4}), (\ref{4.11})\} to find the values of $X$ and $k$.
However, at this point, $T_-$ and $T_+$ are not free parameters. The reason is
that the functions $z(r)$ and $E(r)$ are fully determined by the first-order
equations (\ref{3.18}) and (\ref{3.17}) and by the initial values $z(0) = 0$,
$E(0) = 0$. Consequently, when $z(r)$ is to have the right value at $r_{\rm
AH}$, given by (\ref{2.33}) and (\ref{2.31}), a limitation on $k$ follows. In
fact, $k$ will be determined by trial and error while solving (\ref{3.18}), so
as to ensure that $z(r_{\rm AH}) = z_{\rm AH}$.\footnote{The parameter $k$
enters (\ref{3.18}) via $E$ -- see (\ref{2.21}), and $E$ enters all the
quantities in (\ref{3.6}) -- (\ref{3.18}).} With $k$ given, $T_-$ or $T_+$ are
fixed by (\ref{6.2}) or (\ref{6.4}), and cannot be independently adapted to
observations.

For $k = 0$ we have $X^3 = 2 M_0 H_0$ and
\begin{equation}\label{6.5}
T_0 \df \lim_{X^3 \to 2M_0 H_0} T_- = \frac 2 {3H_0} = 0.099\ {\rm NTU}.
\end{equation}
For $k \to - \infty$ we have $X \to \infty$ and
\begin{equation}\label{6.6}
T_{\infty} \df \lim_{X \to \infty} T_- = \frac 1 {H_0} = 0.149\ {\rm NTU}.
\end{equation}

For the case $k \geq 0$, we use (\ref{6.4}) instead of (\ref{6.2}) to calculate
$T_+$ and obtain
\begin{equation}\label{6.7}
\lim_{X^3 \to 2M_0 H_0} T_+ = \lim_{X^3 \to 2M_0 H_0} T_- = 0.099\
{\rm NTU}.
\end{equation}

See Appendix \ref{Iguchi} for the comparison of the results of this section to
those of Iguchi et al. \cite{INNa2002}.

\section{The equations that determine $z(r)$, ${\cal F}(r)$ and
$E(r)$}\label{numcalzFE}

\setcounter{equation}{0}

To avoid numerical instabilities at $r \to 0$ caused by expressions that become
0/0, we define
\begin{eqnarray}
D_r &\df& {\cal D} / r, \label{7.1} \\
\beta_1 &\df& \frac {B_1} r = \sqrt{- k + {\cal F} + 2 M_0 H_0 (1 + z)/D_r},
\ \ \ \  \label{7.2} \\
\beta_3 &\df& \frac {B_3} {B_1} = \frac 3 2\ \frac{M_0 H_0 (1 + z)} {(- k +
{\cal F})^{3/2}} \nonumber \\
&\times& \left[\sqrt{{\cal U}^2 - 1} - \ln \left({\cal U} + \sqrt{{\cal
U}^2 - 1}\right)\right], \label{7.3} \\
\beta_4 &\df& \frac {B_4} r = \frac{A_1} {\sqrt{1 + 2E} B_2} \left[\frac
{\beta_1} 2 - \frac {M_0 H_0 (1 + z) \beta_3} {{D_r}^2}\right] \nonumber \\
&+& \beta_1 \beta_3 - D_r, \label{7.4}
\end{eqnarray}
and rewrite (\ref{3.20}) and (\ref{3.21}) in the form
\begin{eqnarray}
B_5 &=& 2 \beta_1 \beta_3 - 3 D_r - 2\ \frac {A_1} {B_2}\ \frac {\beta_3}
{{D_r}^2}\ \frac {M_0 H_0 (1 + z)} {\sqrt{1 + 2E}},\ \ \ \ \  \label{7.5} \\
\dr {\cal F} r &=& \frac {- k + {\cal F}} r\ \frac {{\cal A}_1} {{\cal A}_2},
\label{7.6}
\end{eqnarray}
where
\begin{eqnarray}
{\cal A}_1 &\df& - D_r \frac {B_2} {A_1} - \frac {\beta_1} {\sqrt{1 + 2E}},
\label{7.7} \\
{\cal A}_2 &\df& \frac {B_2} {A_1}\ \beta_4 = \frac 1 {\sqrt{1 + 2E}}
\left[\frac {\beta_1} 2 - \frac {M_0 H_0 (1 + z) \beta_3} {{D_r}^2}\right]
\nonumber \\
&+& \frac {B_2} {A_1} \left(\beta_1 \beta_3 - D_r\right), \label{7.8}
\end{eqnarray}
and $B_2$ is in the form (\ref{3.11}). The quantities $\beta_1, \beta_3$ and
$\beta_4$ have well-defined values at $r = 0$, while $D_r$ behaves in a
controllable way at small $r$. The form (\ref{3.11}) of $B_2$ makes the
numerical calculation of the locus of $B_2 = 0$ independent of the precision in
calculating $E(r_{\rm AH})$.

Equation (\ref{3.18}), even with the substitutions listed above, results in a
function $z(r)$ that does not hit the point $(r, z) = (r_{\rm AH}, z_{\rm AH})$
with a satisfactory precision. To improve the precision, (\ref{3.18}) had to be
rewritten as
\begin{equation}\label{7.9}
\dr r z = \frac {2 \beta_4\sqrt{1 + 2E}} {(1 + z) \left[3 {\beta_1}^2 \beta_3 -
3 \beta_1 D_r + (k - {\cal F}) \beta_3\right]}.
\end{equation}
In this form, it was possible to use the tabulated values of the function ${\cal
D}(z)$ -- see Ref. \cite{Kras2014}.

\section{Integration of the set \{(\ref{7.6}), (\ref{7.9})\} for $k >
0$}\label{negativeE}

\setcounter{equation} {0}

The numerical integration of the set \{(\ref{7.6}), (\ref{7.9})\} was first
attempted with $k > 0$. From (\ref{2.21}) and (\ref{2.22}) it then follows that
there is a range $0 < r < r_0$ in which $E  < 0$. As explained in Sec.
\ref{limits0withE}B, handling $E < 0$ requires replacing $B_3$ with the
$\widetilde{B}_3$ given by (\ref{3.26}). Consequently, $\beta_3$ has to be
replaced with
\begin{equation}\label{8.1}
\widetilde{\beta}_3 \df \frac {\widetilde{B}_3} {B_1} = \frac 3 2\ \frac{M_0
H_0 (1 + z)} {(k - {\cal F})^{3/2}} \left(\arccos {\cal U} - \sqrt{1 - {\cal
U}^2}\right).
\end{equation}

\begin{figure}
\hspace{-4cm}
\includegraphics[scale=0.5]{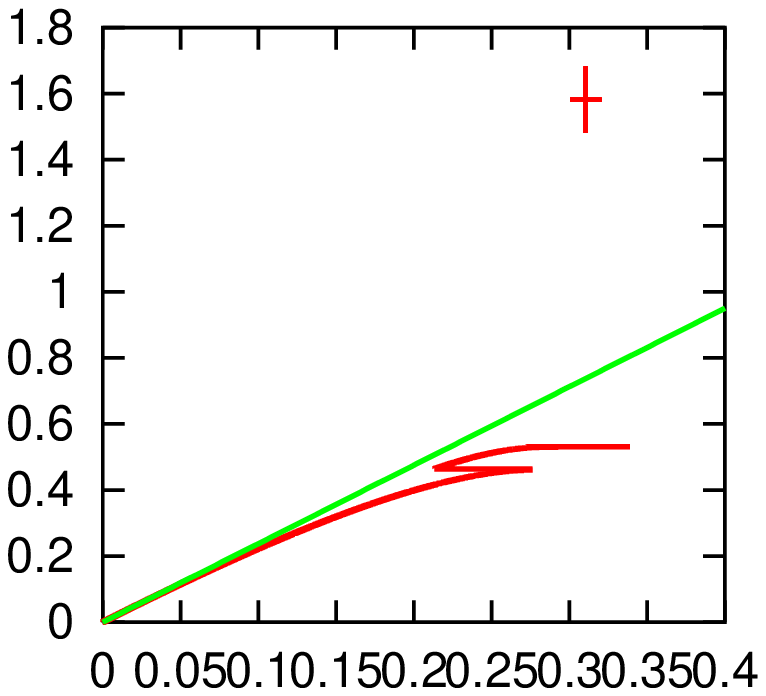}
${ }$ \\[-3.8cm]
\hspace{4cm}
\includegraphics[scale=0.5]{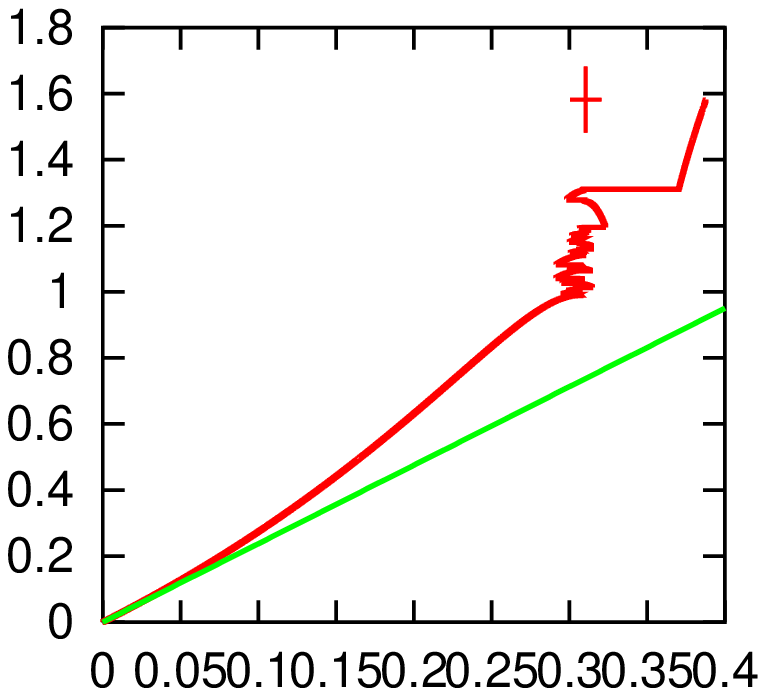}
\caption{Typical graphs of the function $z(r)$ for two ranges of $k > 0$. The
cross marks the point of coordinates $(r, z) = (r, z)_{\rm AH}$, given by
(\ref{2.33}) and (\ref{2.31}). The straight lines are the tangents to $z(r)$ at
$r = 0$, found by solving (\ref{4.11}). {\bf Left panel:} $k = 10^{-3}$. {\bf
Right panel:} $k = 10^{-16}$.} \label{negEfig}
\end{figure}

The functions $r(z)$ and $E(r)$ were calculated for the following values of $k$:
\begin{equation}\label{8.2}
k_j = j, \qquad k_i = 10^{-i},
\end{equation}
with $1 \leq j \leq 10$ and $1 \leq i \leq 16$ being integer. Taking $i > 16$
led to $z(r)$ curves identical to that for $i = 16$. Taking $k = 0$ caused an
immediate breakdown of the calculation -- the limit $k \to 0$ of the formulae is
too tricky for a numerical program. The other results were the following:

For all $1 \leq j \leq 10$, for $i \leq 4$ and $i = 6$, the whole $z(r)$ curve
lies below its tangent at $r = 0$, except for wild numerical fluctuations at the
right end that in some cases go above the tangent. The tangent passes under the
point $(r_{\rm AH}, z_{\rm AH})$ in all these cases. A typical example is the
graph for $i = 3$ shown in the left panel of Fig. \ref{negEfig}. All $z(r)$
curves of this collection end far below $z = z_{\rm AH}$.

For $i = 5$, numerical instabilities kill the calculation already at step 2.

For $i = 7$, the $z(r)$ curve goes off from $r = 0$ very nearly along its
tangent, but the calculation ends in a numerical crash already at step 695, with
$r \approx 0.004592$.

For each $i \geq 8$, the $z(r)$ curve lies above its tangent at $r = 0$, but
goes around the point $(r_{\rm AH}, z_{\rm AH})$ at large distance. With $i \geq
8$, the $z(r)$ curves look similar to each other, except for the shape of the
instabilities at the right end. For $i \geq 16$, even the instabilities have
identical shapes. A typical example of the $i \geq 8$ collection is the graph
for $i = 16$ shown in the right panel of Fig. \ref{negEfig}.

With $E < 0$, the inequality $2E > -1$ must be obeyed at all $r > 0$, see
(\ref{2.3}) and the remark below it. It is obeyed indeed, except at the last
step before the numerical crash, in those cases, where it occurred. The last
value of $E$ yet calculated is $E_l < -1/2$ in all $j$-cases, and with $i = 2,
3, 6$, and going through $E = -1/2$ may have been the reason of the crash. The
exceptions are the cases $i = 4$ and $i = 7$, where the last $E$ is positive,
but these are the end points of wildly fluctuating segments -- and here, going
through $E = 0$ may have been the reason of the final crash. For all $i \geq 8$,
$E$ stays very close to 0, is negative at all $r > 0$, and the calculation does
not crash up to $z_{\rm AH}$, although there are wild fluctuations in both
$z(r)$ and $E(r)$ close to $r = r_{\rm AH}$.

Thus, the conclusion is that the curve $z(r)$ will never hit the point $(r, z) =
(r_{\rm AH}, z_{\rm AH})$ when $k \geq 0$. Consequently, from now on we will
consider only $k < 0$.

\section{Integration of the set \{(\ref{7.6}), (\ref{7.9})\} for $k <
0$}\label{positiveE}

\setcounter{equation} {0}

The best-fit value of $k$ was found experimentally while numerically integrating
the set \{(\ref{7.9}), (\ref{7.6})\}; it is
\begin{equation}\label{9.1}
k = - 21.916458.
\end{equation}
This is the curvature index of the Friedmann model that evolves by the same law
as the central particle in our L--T model. The corresponding $X$ was found from
(\ref{4.11}):
\begin{equation}\label{9.2}
X = 4.961958808006444.
\end{equation}
The age of the Universe in this model is found from (\ref{4.2}) and (\ref{6.2})
to be
\begin{equation}\label{9.3}
T_{\rm model} = 0.1329433206844743\ {\rm NTU} \approx 13.03 \times 10^9\ {\rm
y}.
\end{equation}
Assuming that the vertex of the light cone is at $(t, r) = (0, 0)$, we see from
(\ref{6.2}) and (\ref{9.3}) that
\begin{equation}\label{9.4}
t_B = - T_{\rm model} = - 0.1329433206844743\ {\rm NTU}.
\end{equation}

\begin{figure}
\hspace{-5mm}
\includegraphics{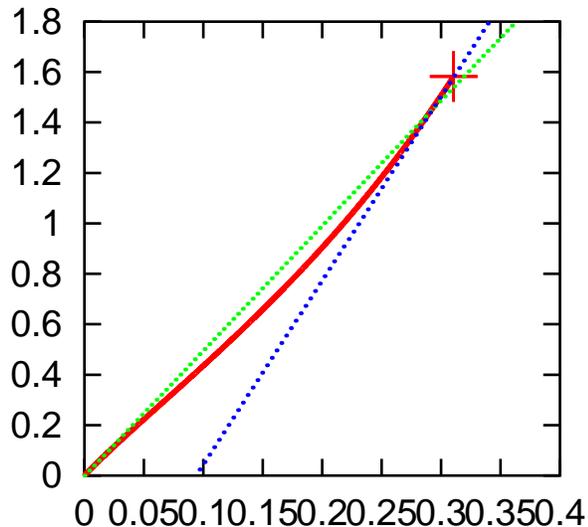}
\caption{Graph of $z(r)$ for $0 \leq r \leq r_{\rm AH}$. The cross marks the
point of coordinates $(r, z) = (r, z)_{\rm AH}$, given by (\ref{2.33}) and
(\ref{2.31}). The dotted straight lines are the tangents to $z(r)$ at $r = 0$
(given by (\ref{9.2})) and at $r = r_{\rm AH}$ (given by (\ref{10.2})).}
\label{zgraph}
\end{figure}

\begin{figure}
\hspace{-5mm}
\includegraphics[scale=0.5]{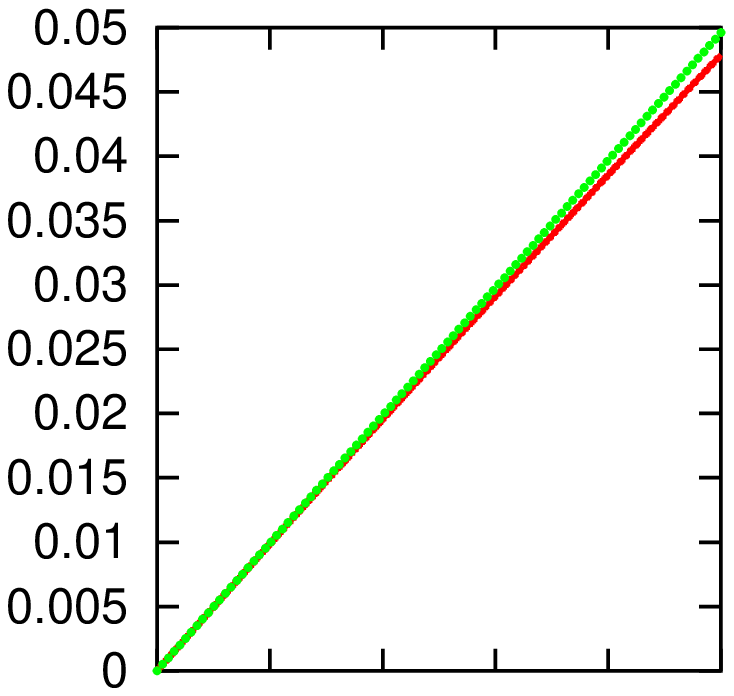}
\hspace{-5mm}
\includegraphics[scale=0.55]{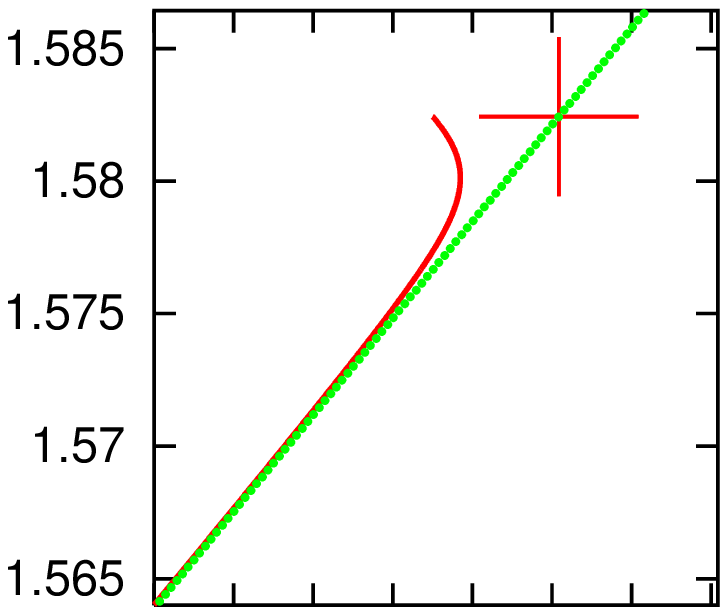}
\caption{{\bf Left panel:} Closeup view of the vicinity of $r = 0$ in Fig.
\ref{zgraph}. The horizontal axis goes from $r = 0$ to $r = 0.01$, the tics on
it are separated by $\Delta r = 0.002$. {\bf Right panel:} Closeup view of the
vicinity of $r = r_{\rm AH}$ in Fig. \ref{zgraph}. The cross marks the point
$(r, z) = (r_{\rm AH}, z_{\rm AH})$, given by (\ref{2.33}) and (\ref{2.31}). The
straight line is the theoretical tangent to $z(r)$ at $r = r_{\rm AH}$ given by
(\ref{10.2}). This mismatch is the best accuracy achieved in Fortran 90 at
double precision. The leftmost tic on the horizontal axis is at $r = 0.3085$,
the rightmost one is at $r = 0.3115$, the tics are separated by $\Delta r =
0.0005$.} \label{zgraphclose}
\end{figure}

\begin{figure}
\hspace{-5mm}
\includegraphics{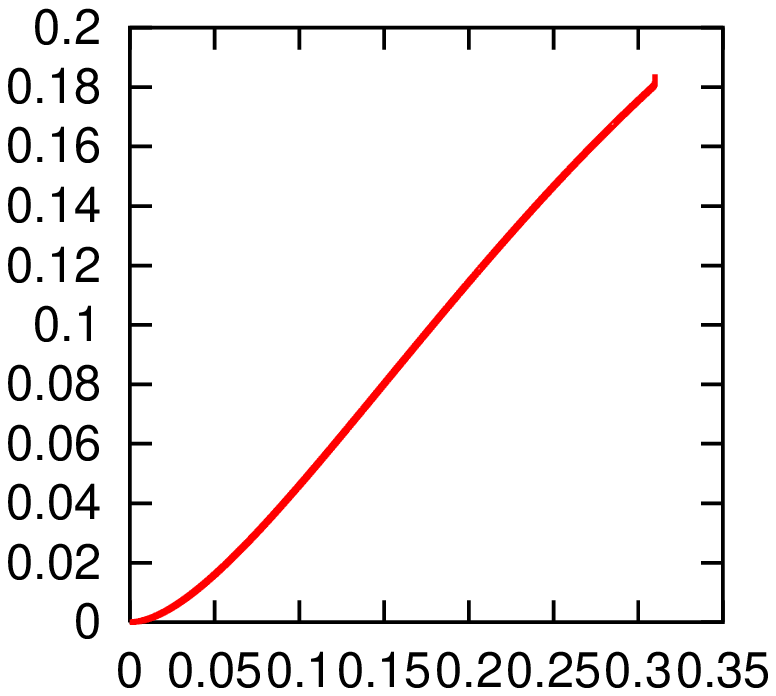}
${ }$ \\[-6.9cm]
\hspace{-2cm}
\includegraphics[scale=0.38]{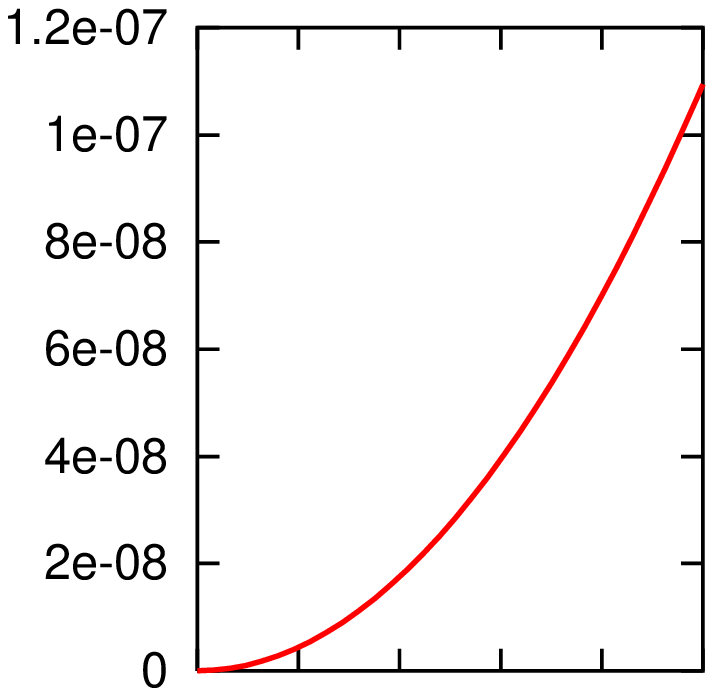}
\vspace{3.6cm} \caption{{\bf Main panel:} Graph of the function $E(r)$ for $0
\leq r \leq r_{\rm AH}$. An instability is seen near $r = r_{\rm AH}$ -- see
Fig. \ref{EgraphatAH}. {\bf Inset:} Closeup view of the vicinity of $r = 0$.
There are no instabilities in this range. The horizontal axis goes from $r = 0$
to $r = 10^{-4}$, the tics are separated by $\Delta r = 2 \times 10^{-5}$.}
\label{Egraph}
\end{figure}

\begin{figure}
\hspace{-5mm}
\includegraphics{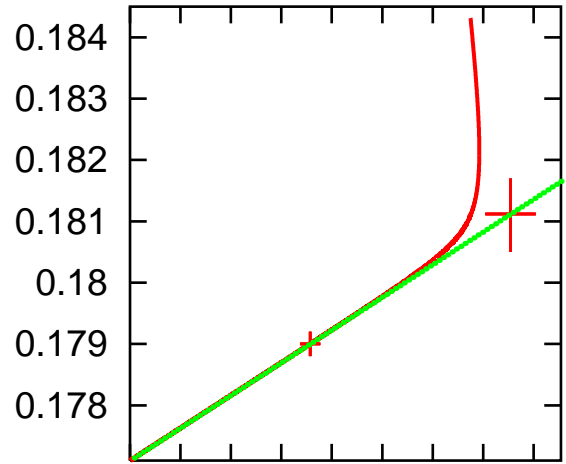}
\caption{Graph of the function $E(r)$ in the vicinity of $r_{\rm AH}$. The
straight line nearly coincides with $E(r)$ for $0.1771 < E < 0.179$. The larger
cross marks the point of coordinates $(r_{\rm AH}, \widetilde{E}_{\rm AH})$,
where $\widetilde{E}_{\rm AH}$ is given by (\ref{9.7}). The leftmost tic on the
horizontal axis is at $r = 0.304$, the rightmost one is at $r = 0.311$, the tics
are separated by $\Delta r = 0.001$. See text for more explanation. }
\label{EgraphatAH}
\end{figure}

Figures \ref{zgraph} and \ref{Egraph} show the results of integration of the set
\{(\ref{7.9}), (\ref{7.6})\} for $r \in [0, r_{\rm AH}]$. Figures
\ref{zgraphclose} and \ref{EgraphatAH} show closeup views of characteristic
regions of the main graphs.

The endpoint of $z(r)$ misses the point $(r_{\rm AH}, z_{\rm AH})$ in Fig.
\ref{zgraphclose} in consequence of numerical errors, but this is the best
precision that could be achieved. Below the order $10^{-6}$, $z(r)$ in the
vicinity of $r_{\rm AH}$ becomes ``quantized'': a change of $k$ at the level of
$10^{-7}$ causes no effect, while a change at the level of $10^{-6}$ causes a
jump of the endpoint that leads to a greater error than the one in the figure.
This happens because, for numerical integration, the segment $[0, r_{\rm AH}]$
was divided into $10^5$ parts -- so $\Delta r \approx 0.31 \times 10^{-5}$ is
the limit of numerical accuracy.

The straight line is the tangent to $z(r)$ at $r = r_{\rm AH}$ given by
(\ref{10.2}). The same numerical errors cause that $z(r)$ does not have the
right slope close to $r = r_{\rm AH}$.

The errors in computing $z(r)$ caused errors in $E(r)$ -- the latter curve also
failed to reach $r = r_{\rm AH}$, as shown in Fig. \ref{EgraphatAH}. But the
precise value of $E$ at $r_{\rm AH}$ must be known in order to calculate the
tangents to $z(r)$ and $E(r)$ at $r_{\rm AH}$, as seen from (\ref{5.3}) --
(\ref{5.6}), which are needed to continue the integration of (\ref{3.18}) and
(\ref{7.6}) beyond $r = r_{\rm AH}$. This difficulty was solved as described
below.

The segment of the $E(r)$ curve in Fig. \ref{EgraphatAH} between the values
$\widetilde{E}_1 = 0.1771$ and $\widetilde{E}_2 = 0.179$ is very nearly
straight. Consequently, it was assumed that it is actually straight. The $r_1$
corresponding to the first $E$ after $\widetilde{E}_1$ (call it $E_1$) and the
$r_2$ corresponding to the first $E$ after $\widetilde{E}_2$ (call it $E_2$)
were read out from the table representing the numerically calculated $E(r)$, and
a straight line was drawn through the points $(r_1, E_1)$ and $(r_2, E_2)$. The
two points are shown in Fig. \ref{EgraphatAH}: the first one coincides with the
lower left corner, the second one is marked with the small cross. Their
coordinates are
\begin{eqnarray}
&& \left(\begin{array}{ll}
r_1 \\
E_1 \\
\end{array}\right) = \left(\begin{array}{ll}
0.3030042702756812 \\
0.1771007383202457 \\
\end{array}\right), \label{9.5} \\
&& \left(\begin{array}{ll}
r_2 \\
E_2 \\
\end{array}\right) = \left(\begin{array}{ll}
0.3065701986604748 \\
0.1790011442990486 \\
\end{array}\right). \label{9.6}
\end{eqnarray}
The intersection of this line with $r = r_{\rm AH}$ occurs at
\begin{equation}\label{9.7}
E = \widetilde{E}_{\rm AH} = 0.18111827859273.
\end{equation}
Since the $E(r)$ curve is as unstable for $r \to r_{\rm AH}$ as Fig.
\ref{EgraphatAH} shows, the construction that led to (\ref{9.7}) could not be
precise. The $\widetilde{E}_{\rm AH}$ of (\ref{9.7}) was taken as the starting
point of the fitting procedure that resulted in the $E(r_{\rm AH}) = E_{\rm AH}$
given by (\ref{10.1}). The point $(r_{\rm AH}, \widetilde{E}_{\rm AH})$ is
marked by the larger cross in Fig. \ref{EgraphatAH}; the corrected point
$(r_{\rm AH}, E_{\rm AH})$ is at this scale indistinguishable from the one
shown.

\section{Verifying the results of Sec. \ref{positiveE}}\label{verifyback}

\setcounter{equation}{0}

\begin{figure}
\hspace{-5mm}
\includegraphics[scale=0.5]{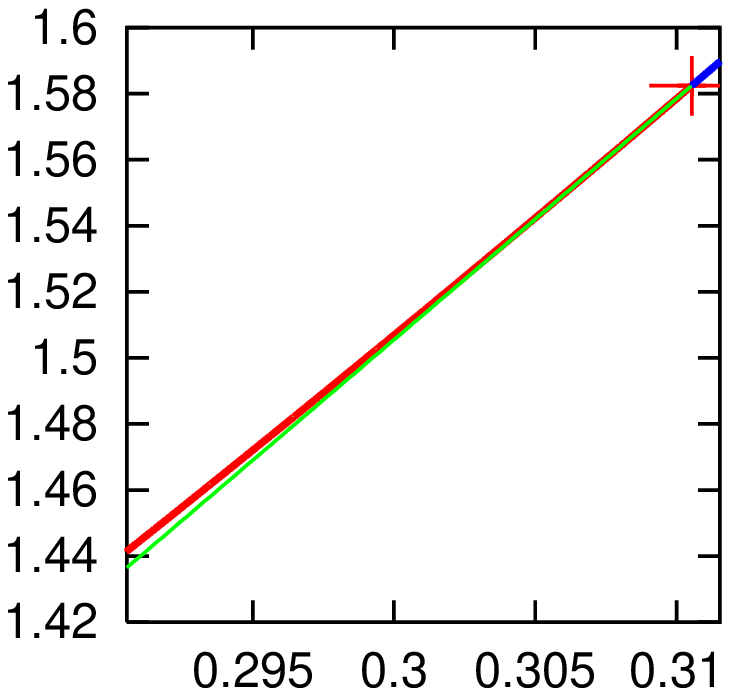}
\hspace{-5mm}
\includegraphics[scale=0.5]{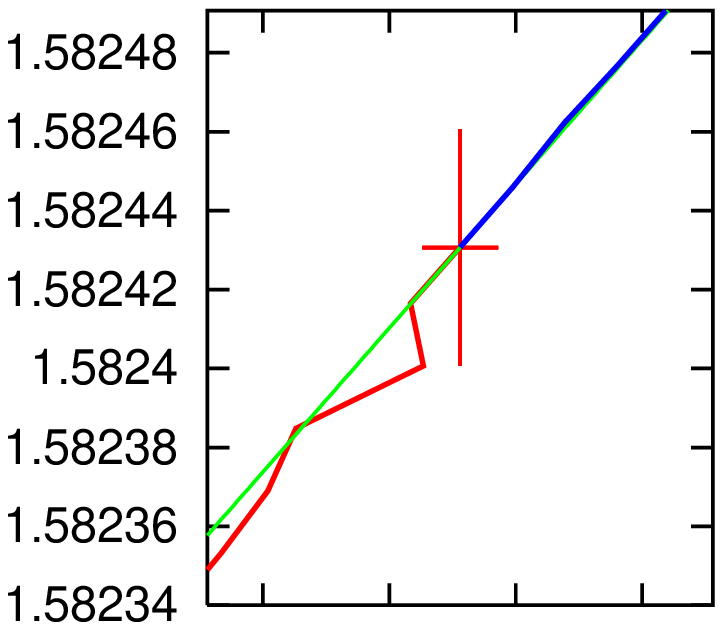}
\caption{{\bf Left panel:} Closeup view of the vicinity of $r = r_{\rm AH}$ on
the curve $z(r)$ obtained by integrating (\ref{7.9}) backward and forward from
the initial point at $r = r_{\rm AH}$ (for information on the forward part see
Sec. \ref{continuepastAH}). The lower line in the left half is the tangent to
$z(r)$ at $r_{\rm AH}$ given by (\ref{10.2}). {\bf Right panel:} A magnified
view of the vicinity of $r = r_{\rm AH}$. The leftmost tic on the horizontal
axis is at $r = 0.310535$, the rightmost one is at $r = 0.31055$, the tics are
separated by $\Delta r = 5 \times 10^{-6}$. The errors in $r$ show up at the
level of $10^{-6}$ in the backward-integrated segment; in the forward-integrated
segment they are a few times smaller. The cross marks the point $(r_{\rm AH},
z_{\rm AH})$.} \label{zbackcloseatAH}
\end{figure}

\begin{figure}
\hspace{-5mm}
\includegraphics[scale=0.5]{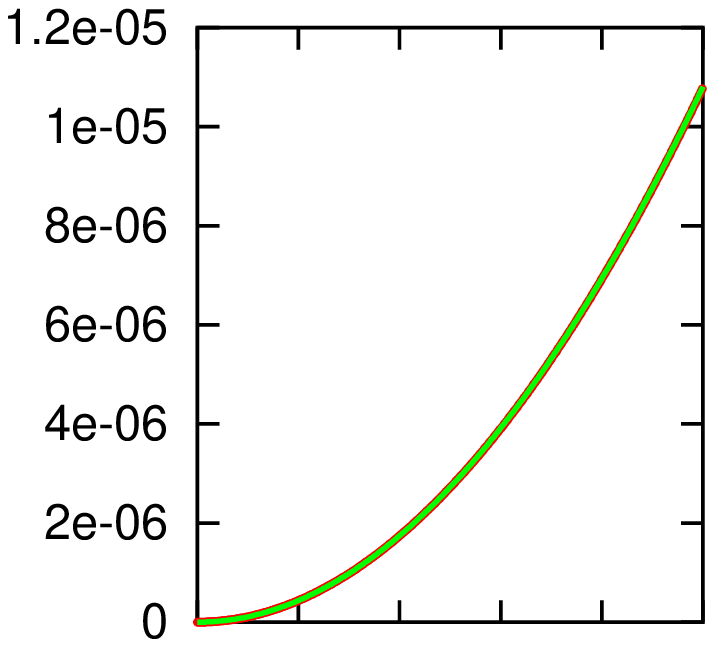}
\hspace{-5mm}
\includegraphics[scale=0.5]{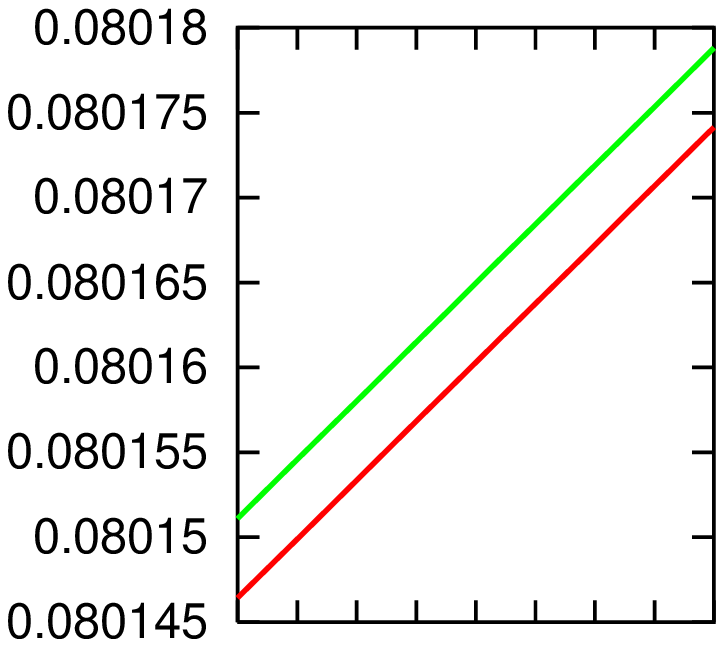}
\caption{A comparison of the two $E(r)$ curves. {\bf Left panel:} In a
neighbourhood of $r = 0$ the backward-integrated $E(r)$ is, at this scale and at
all smaller scales, indistinguishable from the forward-integrated one. The left
margin of the figure is at $r = 0$, the right one at $r = 0.001$, the tics on
the horizontal axis are separated by $\Delta r = 0.0002$. {\bf Right panel:}
Around $r = 0.15$, the two curves differ by $\Delta E = 5 \times 10^{-6}$. The
backward-integrated $E(r)$ is the upper curve. The left margin of the figure is
at $r = 0.14998$, the right margin is at $r = 0.15002$, the tics on the
horizontal axis are separated by $\Delta r = 5 \times 10^{-6}$. }
\label{comparetheEs}
\end{figure}

\begin{figure}
\hspace{-5mm}
\includegraphics[scale=0.53]{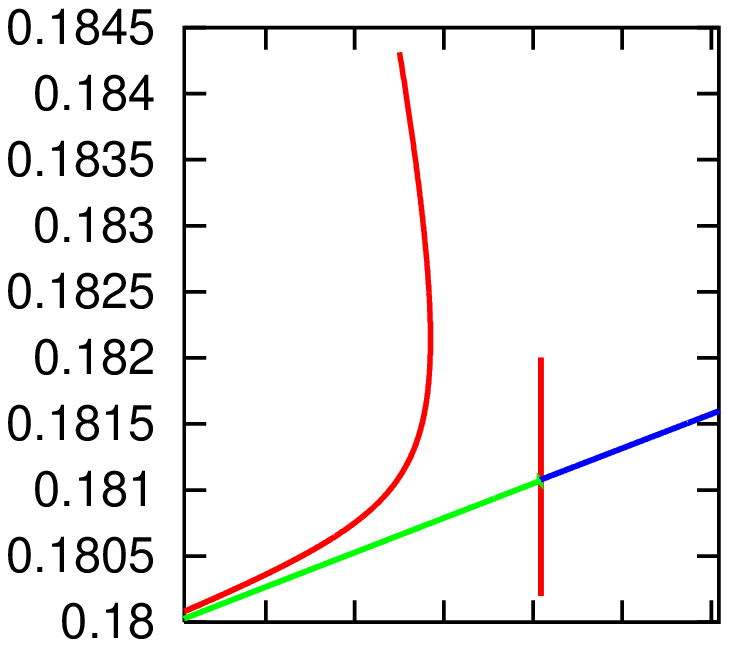}
\hspace{-8mm}
\includegraphics[scale=0.53]{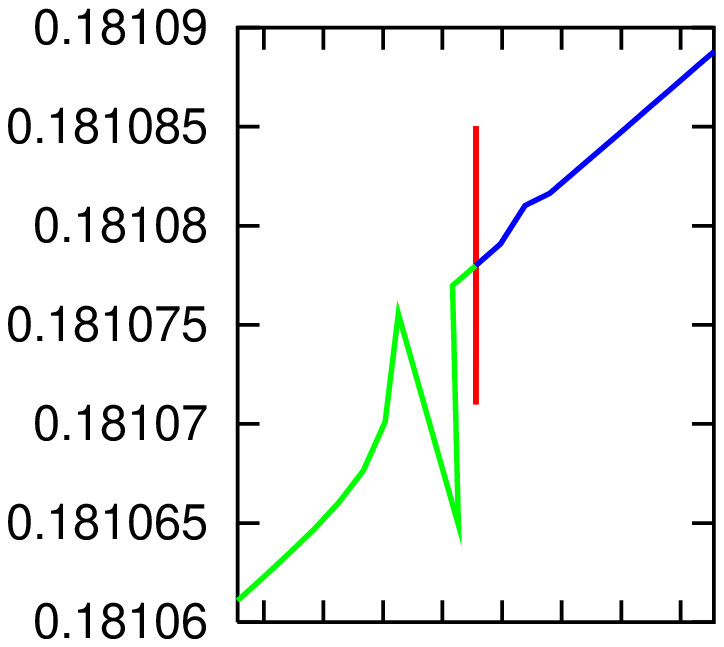}
\caption{The $E(r)$ curve in the neighbourhood of $r = r_{\rm AH}$, marked by
the vertical stroke in both panels. {\bf Left panel:} The curve that bends up is
$E(r)$ integrated forward from $r = 0$. The other line is $E(r)$ integrated
backward and forward from $r = r_{\rm AH}$. The leftmost tic on the horizontal
axis is at $r = 0.309$, the rightmost one is at $0.3115$, the tics are separated
by $\Delta r = 5 \times 10^{-4}$. {\bf Right panel:} The neighbourhood of $r =
r_{\rm AH}$ magnified $\approx 100$ times. The leftmost tic on the horizontal
axis is at $r = 0.310525$, the rightmost one is at $r = 0.31056$, the tics are
separated by $\Delta r = 5 \times 10^{-6}$.} \label{EatAH}
\end{figure}

The computations reported in Sec. \ref{positiveE} were verified by integrating
\ref{7.9}) and (\ref{7.6}) backward from the initial point at $r = r_{\rm
AH}$, with $z_{\rm AH}$ given by (\ref{2.31}). The value of $\widetilde{E}_{\rm
AH}$ given by (\ref{9.7}) was corrected by trial and error so as to ensure that
the curve $E(r)$ integrated backward from $r = r_{\rm AH}$ hits the point $(r,
E) = (0, 0)$ with the maximal precision. The corrected value that emerged is
\begin{equation}\label{10.1}
E_{\rm AH} = 0.181078.
\end{equation}
With $E_{\rm AH}$ now known, we can calculate from (\ref{5.3}) -- (\ref{5.6})
\begin{equation}\label{10.2}
\left(\dr z r\right)_{\rm AH} = 7.29532880561771,
\end{equation}
and from (\ref{5.7}) -- (\ref{5.8}) using (\ref{9.4})
\begin{equation}\label{10.3}
t_{\rm AH} = -0.0966669255756665\ {\rm NTU}.
\end{equation}
With (\ref{10.1}) and (\ref{10.2}), the $z(r)$ and $E(r)$ curves integrated
backward from $r = r_{\rm AH}$ are, at the scale of Figs. \ref{zgraph} and
\ref{Egraph}, indistinguishable from the curves shown there. The precision of
coincidence is shown in Figs. \ref{zbackcloseatAH} -- \ref{EatAH}.

The left panel of Fig. \ref{zbackcloseatAH} is at a scale approx. 10 times
larger than the right panel of Fig. \ref{zgraphclose} and shows a dramatic
improvement of precision -- no instabilities are seen (if the scale were the
same, the $z(r)$ curve would now be indistinguishable from its tangent). The
right panel shows a magnified view of the neighbourhood of $(r, z) = (r_{\rm
AH}, z_{\rm AH})$. The errors in $r$ are seen\footnote{Since the curve $z(r)$
shown in Fig. \ref{zbackcloseatAH} was obtained by integrating (\ref{7.9}), the
solution is in fact the function $r(z)$. Thus, the numerically generated errors
affect $r$, not $z$.} only at the level of $\Delta r = 10^{-6}$. Both panels
include the continuation of $z(r)$ to $r > r_{\rm AH}$, calculated as described
in Sec. \ref{continuepastAH}. Numerical fluctuations are seen in the right panel
both in the backward-integrated segment and in the forward-integrated segment,
where they are a few times smaller, and not, in fact, visible in the figure.

Close to $r = 0$, the curves calculated in the two ways are indistinguishable
even at the smallest scales. In the segment around $r = 0.15$, they differ by
$\Delta z \approx 4 \times 10^{-6}$.

Figures \ref{comparetheEs} and \ref{EatAH} show a comparison of the $E(r)$
curves calculated in the two ways.

Figure \ref{EatAH} shows closeup views of the function $E(r)$ in the
neighbourhood of $r = r_{\rm AH}$ at two scales. The curve found by integrating
(\ref{7.6}) forward from $r = 0$ goes off the right course already at $r \approx
0.309$ and does not reach $r_{\rm AH}$. The curve found by integrating
(\ref{7.6}) forward and backward from $r = r_{\rm AH}$ seems to be smooth at
this scale. The right panel shows the neighbourhood of $r = r_{\rm AH}$
magnified $\approx 100$ times with respect to the left panel. At this scale,
fluctuations in the backward-integrated curve are $\Delta E \approx 10^{-5}$,
those in the forward-integrated curve are at the level of $10^{-6}$.

\section{Continuing the integration of (\ref{3.18}) and (\ref{3.21}) beyond the
AH}\label{continuepastAH}

\setcounter{equation}{0}

\begin{figure}
\hspace{-5mm}
\includegraphics{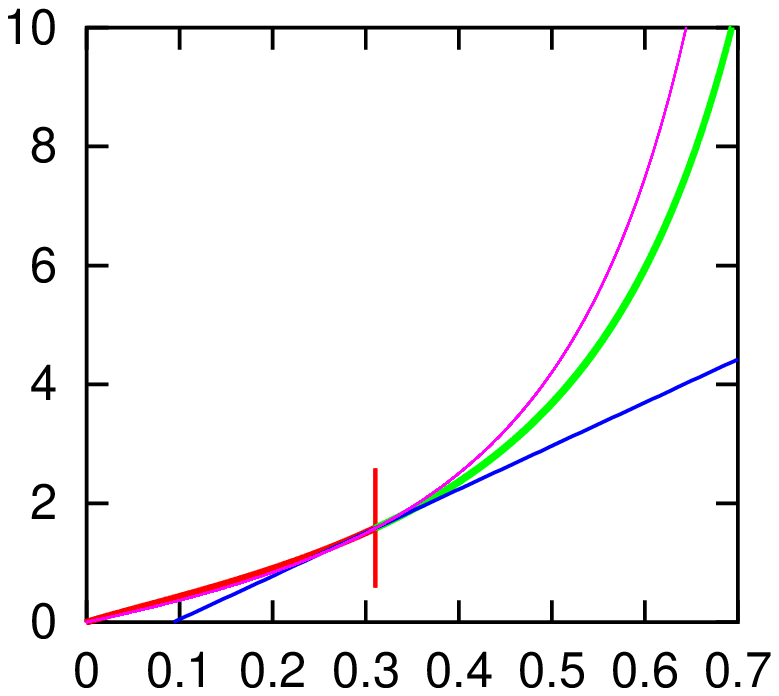}
${ }$ \\[-6.8cm]
\hspace{-1.8cm}
\includegraphics[scale=0.5]{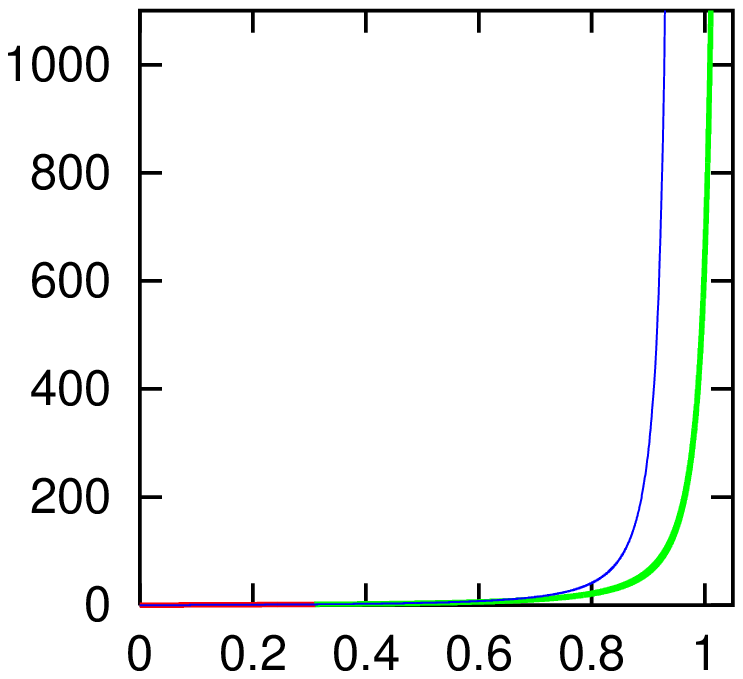}
\vspace{3cm}
\caption{{\bf Main panel:} The continuous curve is the graph of $z(r)$ for $z
\in [0,10]$. The dotted curve is $z(r)$ for the $\Lambda$CDM model -- see text
for explanations. The straight line is the tangent to $z(r)$ at $r = r_{\rm
AH}$, the vertical stroke marks $r = r_{\rm AH}$. {\bf Inset:} The graph of
$z(r)$ for $z \in [0,1100]$. The curve at right is for the L--T model, the curve
at left is for the $\Lambda$CDM model. }
\label{drawzfullrange}
\end{figure}

\begin{figure}
\includegraphics[scale=0.6]{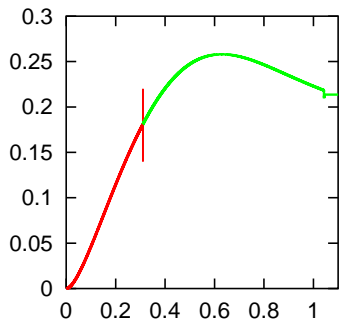}
\caption{The function $E(r)$ extended into the range $r > r_{\rm AH}$. The
vertical stroke is at $r = r_{\rm AH}$. Since $E(r)$ becomes decreasing at $r =
r_{\rm sc}$ given by (\ref{11.3}), there are shell crossings in the region $r >
r_{\rm sc}$. } \label{Efullrange}
\end{figure}

Since by integrating backward from $r = r_{\rm AH}$ (see Sec. \ref{verifyback})
the functions $z(r)$ and $E(r)$ behave controllably in a neighbourhood of the
AH, the calculation of these functions into the range $r > r_{\rm AH}$ could be
undertaken. The independent variable was $r$ and the step in $r$ was $\Delta r =
r_{\rm AH}/(1.5 \times 10^5)$. The corrected value of $E_{\rm AH}$ given by
(\ref{10.1}) was used in all computations and graphs. Figures
\ref{drawzfullrange} -- \ref{Efullrange} show the results (pieces of those
graphs have already been used in Figs. \ref{zbackcloseatAH} and \ref{EatAH}).

The thicker curves in Fig. \ref{drawzfullrange} are the graphs of $z(r)$. The
calculation went up to $r = r_{\rm max}$, achieved at step $n = 355,012$ beyond
$r_{\rm AH}$, with $z = z_{\rm max}$, where
\begin{equation}\label{11.1}
\left(\begin{array}{ll}
r_{\rm max} \\
z_{\rm max} \\
\end{array}\right) = \left(\begin{array}{ll}
1.045516839812362 \\
9.1148372886058313 \times 10^{225} \\
\end{array}\right).
\end{equation}
Then $z$ became too large to handle by Fortran. The main panel in Fig.
\ref{drawzfullrange} shows the range $z \in [0,10]$, the inset shows the range
$z \in [0, 1100]$. The endpoint of this range corresponds to the redshift at
last scattering, which is \cite{Luci2004}
\begin{equation}\label{11.2}
z_{\rm ls} \approx 1089.
\end{equation}
The $r_{\rm max}$ is the approximate value of $r$, at which the past light cone
of the observer reaches the BB set.

This behaviour at approaching the BB is similar to that found in Ref.
\cite{Kras2014}. There, the maximal value of $z$ was $1.6236973619875722 \times
10^{229}$.

The thinner curves in Fig. \ref{drawzfullrange} are the graphs of $z(r)$ for the
$\Lambda$CDM model. There is a subtle point about comparing the $\Lambda$CDM and
L--T models, namely, the $r$-coordinates in them have to be made compatible.
This point was not handled correctly in Ref. \cite{Kras2014}; it is explained in
Appendix \ref{correctr}. As shown there, when the $r$-coordinates are
compatible, the $(r, z)$-coordinates of the AH must be the same in both models.
Indeed, the two graphs of $z(r)$ in Fig. \ref{drawzfullrange} intersect at $(r,
z) = (r_{\rm AH}, z_{\rm AH})$ to better than $10^{-6}$ in each direction. At
all $r < r_{\rm AH}$, the $z(r)$ is smaller in the $\Lambda$CDM model, at $r
> r_{\rm AH}$, the $z(r)$ is larger in the $\Lambda$CDM model. The BB in
the $\Lambda$CDM model, as seen from the inset, corresponds to smaller $r$.

Figure \ref{Efullrange} shows the function $E(r)$ extended into the range $r >
r_{\rm AH}$. It is increasing up to
\begin{equation}\label{11.3}
r = r_{\rm sc} = 0.6293128978680214
\end{equation}
and then begins to decrease. Hence, there are shell crossings in the region $r >
r_{\rm sc}$, see (\ref{2.20}). The redshift corresponding to $r_{\rm sc}$ is
$z_{\rm sc} = 6.938073260172738$. For comparison, the two original projects
investigated supernovae of type Ia having redshifts in the range $0.16 \leq z
\leq 0.62$ \cite{Ries1998} and $0.18 \leq z \leq 0.83$ \cite{Perl1999}, and the
recently discovered most distant Ia supernova has redshift $z = 1.914$
\cite{Jone2014}. Hence, to do away with the shell crossing, our model should be
matched to a background (Friedmann, for example) at $r$ corresponding to the
redshifts in the range $1.914 < z < z_{\rm sc}$, i.e. $0.3525778644179596 < r <
r_{\rm sc}$, and this will not compromise its applicability to the type Ia
supernovae observations.

\section{Calculating the past light cone of the central
observer}\label{lightcone}

\setcounter{equation}{0} \setcounter{table}{0}

At this point, all data needed to numerically solve (\ref{2.9}) are available.
Curiously, the solution turned out to be extremely sensitive to changes of the
algebraic form of the data. For example, a different $t(r)$ curve resulted when
(\ref{3.13}) was combined with (\ref{3.14}) to produce
\begin{equation}\label{12.1}
\dr t r = \frac {A_1} {B_2 H_0 (1 + z)^2 \sqrt{1 + 2E}}\ \dr z r,
\end{equation}
and then $\dril z r$ was replaced by (\ref{3.18}), and still a different curve
when (\ref{3.17}) and (\ref{3.21}) were used in (\ref{3.14}) to eliminate
$(\dril r E)/E$, and the result reparametrised by (\ref{7.1}) -- (\ref{7.3}), to
produce
\begin{eqnarray}\label{12.2}
\dr t r &=& \frac {W_1} {W_2}, \qquad {\rm where} \nonumber \\
W_1 &\df& \frac {r \left(D_r - \beta_1 \beta_3\right)} {- k + {\cal F}}\ \dr
{\cal F} r - D_r, \nonumber \\
W_2 &\df& H_0 (1 + z) \sqrt{1 + 2E}.
\end{eqnarray}
When (\ref{12.1}) was applied in the range $r > r_{\rm AH}$, the curve $z(r)$
failed to reach the BB time given by (\ref{9.4}).

\begin{figure}[h]
\hspace{-11mm}
\includegraphics[scale=0.9]{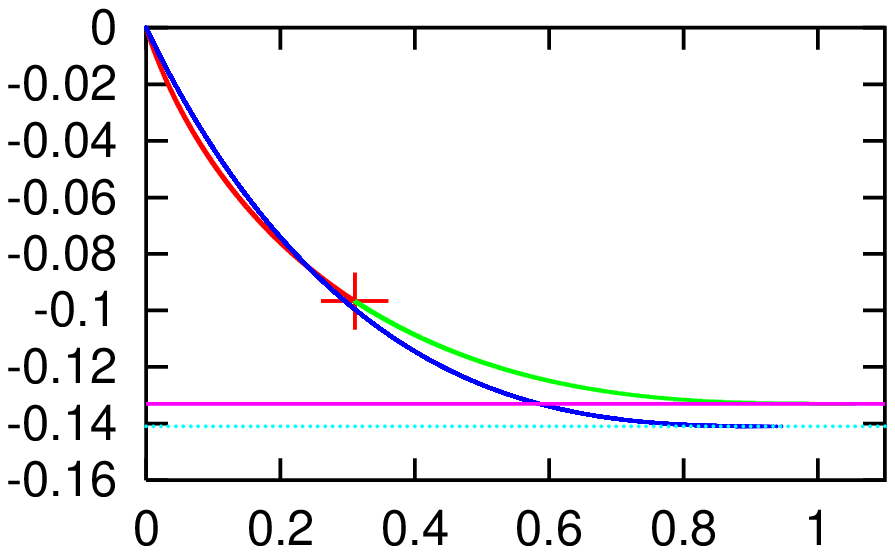}
${ }$ \\[-5cm]
\hspace{8cm}
\includegraphics[scale=0.4]{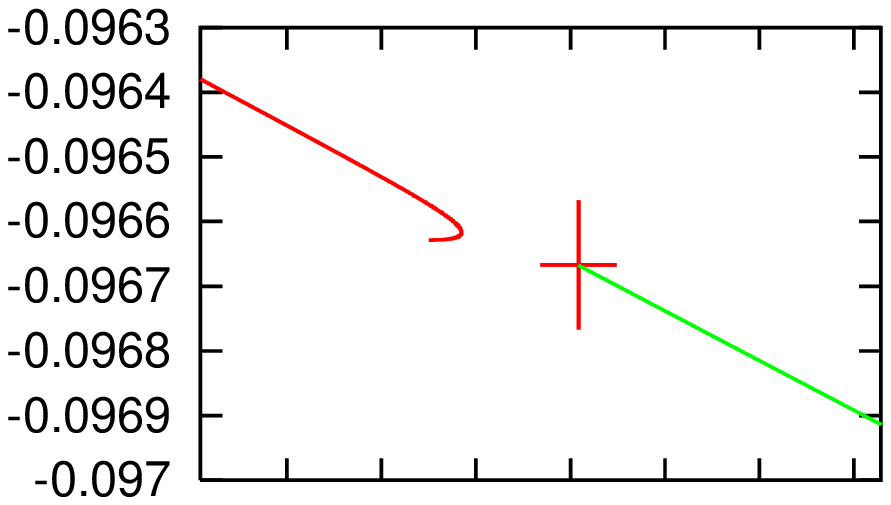}
\vspace{2cm} \caption{The light cone profile obtained by integrating
(\ref{12.1}) from $r = 0$ to $r = r_{\rm AH}$ and then integrating (\ref{12.2})
beyond $r_{\rm AH}$ with the initial values of $r = r_{\rm AH}$ and $t = t_{\rm
AH}$ given by (\ref{2.33}) and (\ref{10.3}). The upper horizontal line is the $t
= t_B$ given by (\ref{9.4}). The cross marks the point $(r_{\rm AH}, t_{\rm
AH})$. The dotted lines are the $\Lambda$CDM light cone and the $\Lambda$CDM
bang time. {\bf Inset:} Closeup view of the neighbourhood of $r = r_{\rm AH}$.
The gap in the $t(r)$ curve is $\delta r \approx 0.0005$. The tics on the
horizontal axis are separated by $\Delta r = 0.0005$, the leftmost one is at $r
= 0.309$, the rightmost one is at $r = 0.312$. } \label{conefirst}
\end{figure}

The most reliable results were obtained when (\ref{12.1}) was used for the
integration from $r = 0$ to $r = r_{\rm AH}$, and (\ref{12.2}) was used for
integration from $r = r_{\rm AH}$ both ways. These results are presented in Fig.
\ref{conefirst}. The $t(r)$ curve found by integrating (\ref{12.1}) forward from
$r = 0$ failed to reach the point with the coordinates $(r_{\rm AH}, t_{\rm
AH})$ given by (\ref{2.33}) and (\ref{10.3}). The gap $\delta r \approx 0.0005$
is invisible at the scale of the main figure; it is shown in the inset. The
dotted lines in Fig. \ref{conefirst} are the $\Lambda$CDM light cone found by
integrating $\dril t r = - S(t)$, with $S$ given by (\ref{2.14}), and the
$\Lambda$CDM Big Bang time given by (\ref{2.39}). The same subtle point about
comparing the $\Lambda$CDM and L--T models that was mentioned below
(\ref{11.3}) has to be observed also here; see Appendix \ref{correctr}.

As seen from the graphs, the $\Lambda$CDM model universe is older than its L--T
counterpart considered here. For the qualitative description of mimicking the
accelerated expansion in the L--T model see Sec. \ref{conclu}.

\begin{figure}[h]
\hspace{-5mm}
\includegraphics[scale=0.4]{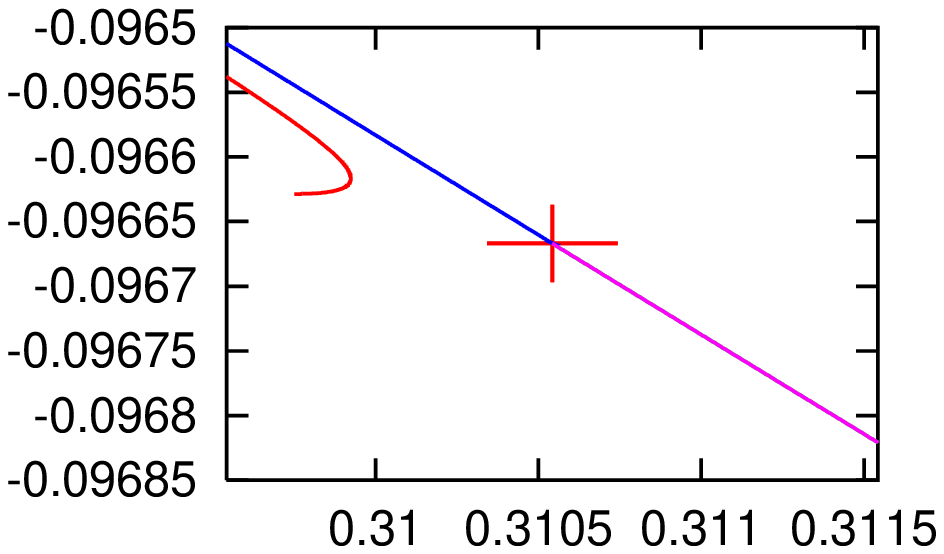}
\includegraphics[scale=0.4]{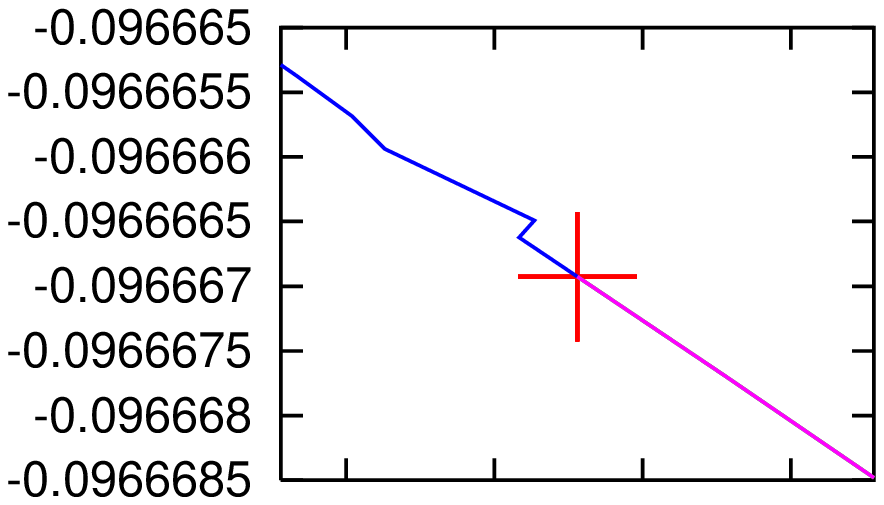}
${ }$ \\[5mm]
\hspace{-5mm}
\includegraphics[scale=0.4]{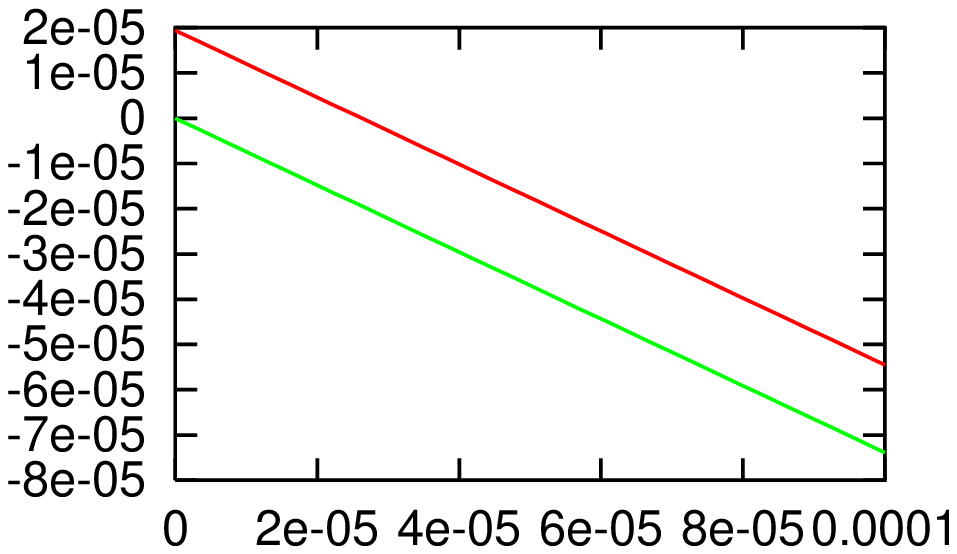}
\includegraphics[scale=0.4]{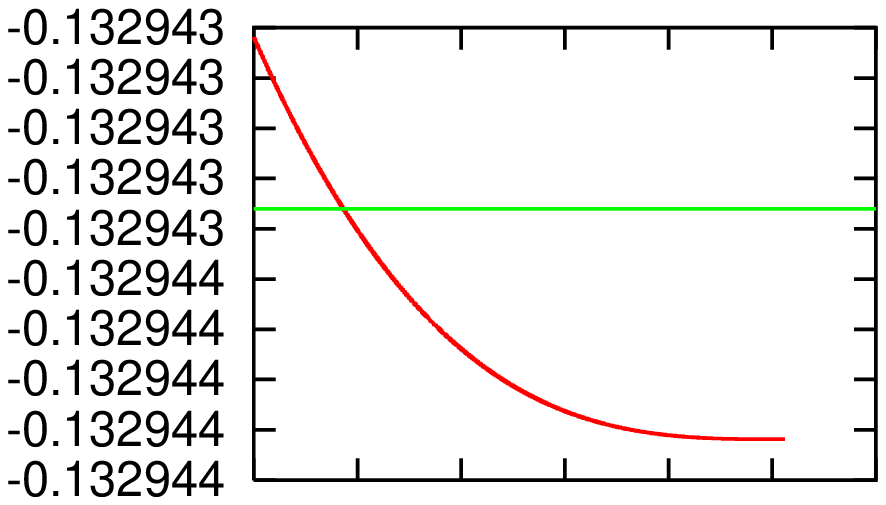}
\caption{Closeup views of key segments of $t(r)$ integrated in the two ways
described in the text. The cross marks the point $(r_{\rm AH}, t_{\rm AH})$.
{\bf Upper left panel:} The neighbourhood of $(r, t) = (r_{\rm AH}, t_{\rm
AH})$. The right end of $t(r)$ integrated forward from $r = 0$ is seen at left.
{\bf Upper right panel:} The neighbourhood of $(r, t) = (r_{\rm AH}, t_{\rm
AH})$ magnified 100 times with respect to the left panel. The leftmost tic on
the horizontal axis is at $r = 0.310535$, the rightmost one is at $r = 0.31055$,
the tics are separated by $\Delta r = 5 \times 10^{-6}$. {\bf Lower left panel:}
The neighbourhood of $r = 0$. The forward-integrated $t(r)$ is the lower curve.
The two curves differ by $\Delta t_1 \approx 1.9 \times 10^{-5}$ NTU $\approx
1.86 \times 10^5$ y. {\bf Lower right panel:} The endpoint of $t(r)$ misses the
BB time by $\Delta t_2 \approx 10^{-6}$ NTU $\approx 9.8 \times 10^4$ y. The
left end of the horizontal axis is at $r = 1.02$, the right end is at $r =
1.05$, the tics are separated by $\Delta r = 0.005$. } \label{alldetails}
\end{figure}

The light cone $t(r)$ integrated backward and forward from the initial point
$(r, t) = (r_{\rm AH}, t_{\rm AH})$, at the scale of the main graph in Fig.
\ref{conefirst}, coincides with the curve shown. Detailed comparisons of the
results of the two integrations are shown in Fig. \ref{alldetails}. The
backward-integrated $t(r)$ misses the point $(r, t) = (0, 0)$ by
\begin{equation}\label{12.3}
\Delta t_1 \approx 1.9 \times 10^{-5}\ {\rm NTU}\ \approx 1.86 \times 10^5\ {\rm
y}.
\end{equation}
At $r = r_{\rm AH}$, the backward branch goes off with fluctuations in $\dril t
r$ caused by jumps in $r(z)$. These could be reduced by increasing the number of
grid points above the current $10^5$. The $t(r)$ curve overshoots the BB by
\begin{equation}\label{12.4}
\Delta t_2 \approx 10^{-6}\ {\rm NTU}\ \approx 9.8 \times 10^4\ {\rm y}.
\end{equation}

Now comes the final test of precision of our numerical calculations. The
right-hand side of (\ref{2.26})
\begin{equation}\label{12.5}
F_r(r) \df {\cal D}/[H_0 (1 + z)]
\end{equation}
comes directly from the input data, via (\ref{2.27}). The left-hand side of
(\ref{2.26})
\begin{equation}\label{12.6}
F_l(r) \df R(t_{\rm ng}(r), r),
\end{equation}
results from the chain of numerical calculations performed in order to find
$z(r)$, $E(r)$ and $t_{\rm ng}(r)$ before $R(t_{\rm ng}(r), r)$ is calculated.
By (\ref{2.26}), the two functions should be identical, so the difference
between them is a measure of precision of the calculation.

\begin{figure}[h]
\includegraphics[scale=0.85]{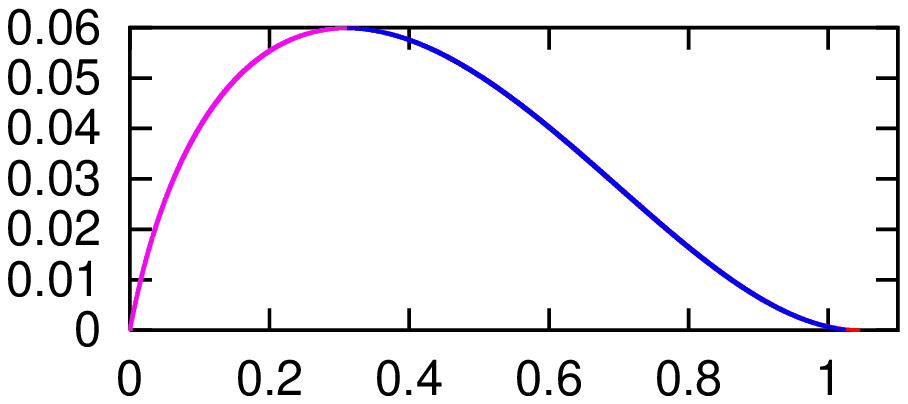}
${ }$ \\[-27mm]
\hspace{-8mm}
\includegraphics[scale=0.35]{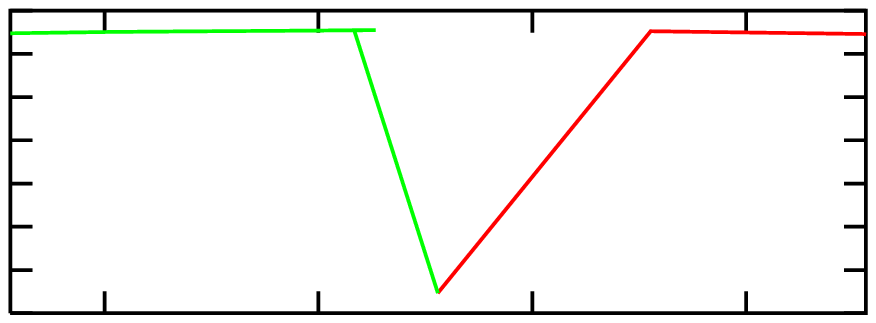}
${ }$ \\[10mm]
\hspace{-3mm}
\includegraphics[scale=0.8]{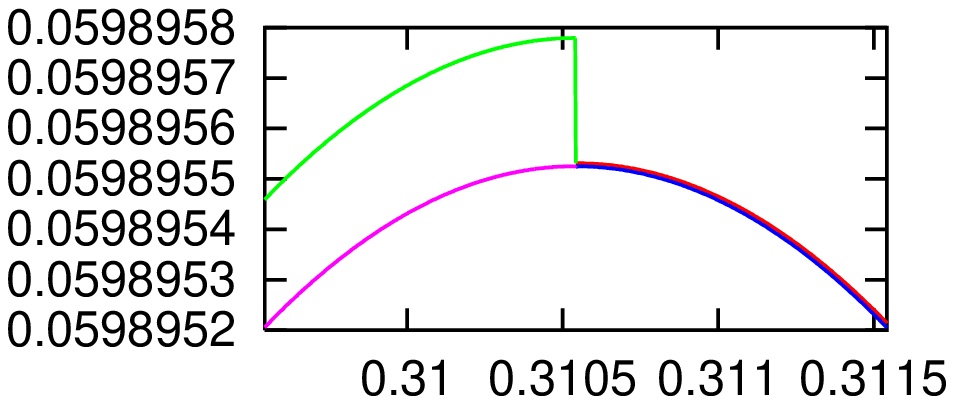}
\caption{Comparison of the function $R(t(r), r)$ along the past light cone
calculated from (\ref{2.26}) with the same function calculated by substituting
$t(r)$, the solution of (\ref{12.2}), into $R(t, r)$. At the scale of the upper
panel, the two functions seem to coincide. The differences between them are
listed in Table \ref{inaccuracies}. The lower panel shows the neighbourhood of
$r = r_{\rm AH}$ -- see the text for an explanation. The inset in the upper
panel shows the numerical instability in $F_l$ at the AH described in the
text. The tics on the horizontal axis are separated by $\Delta r = 5 \times
10^{-6}$, the depth of the dip is $\Delta F_l = 5 \times 10^{-9}$.}
\label{compareRandD}
\end{figure}

Figure \ref{compareRandD} shows the comparison of $F_r(r)$ with $F_l(r)$,
calculated backward and forward from the initial point at $r = r_{\rm AH}$. At
the scale of the upper panel of Fig. \ref{compareRandD}, the two curves are
indistinguishable, but closeup views (not shown) reveal the differences listed
in Table \ref{inaccuracies}.

\begin{table}[h]
\begin{center}
\caption{Discrepancies between (\ref{12.5}) and
(\ref{12.6})}\label{inaccuracies}
\begin{tabular}{|l|l|}
 \hline \hline
At $r =$ & the difference between the two curves is \\
 \hline \hline
0, $2 \times 10^{-5}$ & 0 (invisible for Gnuplot at scales \\
and 0.15 & down to $\Delta r = 10^{-6}$) \\
 \hline
0.25 & $2 \times 10^{-7}$ \\
 \hline
0.6 & $1.97 \times 10^{-6}$ \\
 \hline
1 & $5 \times 10^{-5}$ \\
 \hline \hline
\end{tabular}
\end{center}
\end{table}

The lower panel in Fig. \ref{compareRandD} shows the more complicated situation
in the vicinity of $r = r_{\rm AH}$. The upper curves on both sides of the jump
are the $F_r(r)$, the other curves are the $F_l(r)$. The jump $\Delta F_r = 2.45
\times 10^{-7}$ at the AH is a consequence of the way in which ${\cal D}_{\rm
AH}$ was calculated and ${\cal D}$ tabulated.\footnote{See Ref. \cite{Kras2014}
for a description. In brief, an upper bound $Z > z_{\rm AH}$ was first estimated
approximately, and then the interval $[0, Z]$ was divided into $10^9$ segments
in order to calculate $z_{\rm AH}$ and ${\cal D}_{\rm AH}$ exactly. However,
using $10^9$ points for each of the many calculations would make the progress
prohibitively slow. So, the table of values of ${\cal D}(z)$ for $z \in [0,
z_{\rm AH}]$ was calculated only for $10^5$ intermediate points. The cumulative
numerical error caused the jump between the $(10^5 - 1)$st value of ${\cal D}$
and ${\cal D}_{\rm AH}$, of the order of $\Delta {\cal D} \approx 10^{-6}$; its
consequences are seen in Fig. \ref{compareRandD}.} There is a numerical
instability on each side of the AH that caused a fluctuation in $F_l$ of the
order of $5 \times 10^{-9}$ in the first step of integration. However, at the
second step, the two branches of $F_l$ have the same value on both sides of the
AH down to scales smaller than $10^{-9}$. The difference between $F_r$ and $F_l$
is $6.25 \times 10^{-9}$ for $r > r_{\rm AH}$ and $ 2.45 \times 10^{-7}$ for $r
< r_{\rm AH}$. This precision could be improved by increasing the number of grid
points above the $10^5$ used throughout this paper.

The consistency between $F_l(r)$ and  $F_r(r)$ is somewhat worse if we take the
$t(r)$ integrated forward from $r = 0$ as the basis. Then the two curves agree
perfectly at $r = 0$, but at $r = r_{\rm AH}$ they differ by $2.5 \times
10^{-5}$.

\section{Conclusions}\label{conclu}

\setcounter{equation}{0}

\begin{figure}[h]
\hspace{-5mm}
\includegraphics[scale=0.7]{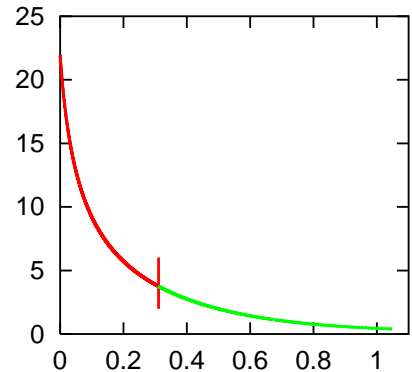}
\caption{Graph of the function $|k_F(r)| \equiv |- k + {\cal F}|$. It is
decreasing all the way to that $r$, at which the past light cone reaches the BB.
Thus, shells of matter closer to the observer evolve by the Friedmann equation
corresponding to larger $|k_F|$. See text for the interpretation. The vertical
stroke is at $r = r_{\rm AH}$. } \label{kequiv}
\end{figure}

Since the function $E(r)$ calculated here generates the same relation $D_L(z)$
as that found in the $\Lambda$CDM model, it imitates the accelerated expansion.
Here is a descriptive explanation of how it happens. The Friedmann limit of our
model is achieved when $-2 E/r^2 \df k_F$ is constant, as stated under
(\ref{2.11}). This $k_F$ is the curvature index of the limiting Friedmann
metric. Since $E/r^2$ is not constant, the $k_F$ will be different at every $r$.
This means that the evolution of each $r =$ constant shell of matter in the L--T
model coincides with the evolution of a different Friedmann model. Figure
\ref{kequiv} shows the function $|k_F(r)| \equiv |- k + {\cal F}|$. It is
decreasing all the way to that $r$, at which the light cone touches the BB set.
Thus, shells of matter closer to the observer evolve by a Friedmann equation
corresponding to larger $|k|$. Consequently, they are ejected from the BB with a
larger value of $\dril S t$ than farther shells, and so intersect the observer's
past light cone with a larger velocity than a Friedmann shell would. Thus,
accelerated expansion is imitated -- without introducing ``dark energy'' or any
other exotic matter.

Recall that the L--T model duplicating the $\Lambda$CDM $D_L(z)$ that was
obtained in Ref. \cite{Kras2014} was rather exceptional: the present observer's
past light cone was the first one that had an infinite redshift at the
intersection with the BB set. All earlier light cones of the central observer
had an infinite blueshift at BB. In the L--T model presented here, all past
light cones of the central observer have infinite redshift at the BB because
$t_B =$ constant. The present model necessarily has shell crossings in the
region $r > r_{\rm sc}$, where $r_{\rm sc}$ is given by (\ref{11.3}). However,
the $r \geq r_{\rm sc}$ region can be cut out of the manifold by matching the
L--T model to a Friedmann background, and this will not harm the applicability
of our model to the Ia supernovae observations, see the final remark in Sec.
\ref{continuepastAH}.

The shell crossings are not necessarily present when both $t_B(r)$ and $E(r)$
are allowed to have non-Friedmannian forms. Examples are the configurations
considered in Ref. \cite{CBKr2010}.

In the L--T model with $E/r^2 =$ constant and variable $t_B$, considered in Ref.
\cite{Kras2014}, the differential equation defining $z(r)$ was uncoupled from
the one that defines $t_B(r)$, so it could be integrated independently. In the
present paper, the equations defining $z(r)$ and $E(r)$, (\ref{7.9}) and
(\ref{7.6}) with (\ref{2.21}), are coupled and have to be integrated
simultaneously. This had no pronounced influence on the precision in calculating
the light cone -- see (\ref{12.3}) and (\ref{12.4}), and the test $F_l(r) =
F_r(r)$ shown in Fig. \ref{compareRandD} came out even better than the one in
Ref. \cite{Kras2014}. The precision could be further improved by increasing the
number of grid points above the $10^5$ used in all programs here.

The present paper revealed the details of geometry of the L--T model that
imitates accelerated expansion of the Universe using $E(r)$ alone, and the
relation of its light cone to that of the $\Lambda$CDM model. It is
complementary to Ref. \cite{Kras2014}, where the same was done for imitating
accelerated expansion with $t_B(r)$ alone. The two papers together are an
extension and complement to Ref. \cite{INNa2002}, in which only a numerical
proof of existence of such L--T models was given. Moreover, in Ref.
\cite{INNa2002}, the numerical integration of the equations corresponding to our
(\ref{2.9}), (\ref{2.10}) and (\ref{2.27}) was carried out only up to the AH,
where the numerics broke down. Consequently, those authors had no chance to
discover the shell crossing because $r_{\rm sc} > r_{\rm AH}$, see (\ref{11.3})
and (\ref{2.33}).

As was shown in Sec. \ref{positiveE}, the value of $k = \lim_{r \to 0}
(-2E/r^2)$ is fixed by the requirement that the $z(r)$ curve passes through the
points $(r, z) = (0, 0)$ and $(r, z) = (r_{\rm AH}, z_{\rm AH})$. The values of
$r_{\rm AH}$ and $z_{\rm AH}$ are, in turn, fixed by the values $H_0$,
$\Omega_m$ and $\Omega_{\Lambda}$, as seen from (\ref{2.27}) -- (\ref{2.30}).
These are taken from observations \cite{Plan2013}. Consequently, it is not
correct to treat $k$ as a free parameter to be determined by observations.
Unfortunately, this conclusion seems to have been unknown to other authors --
exactly this approach was applied in Refs. \cite{INNa2002} and \cite{RCCh2014};
the latter considered a problem equivalent to the present paper by a different
method. The value of $k$ given by our (\ref{9.1}) is not in the collection
considered in the two papers. See Appendices \ref{Iguchi} and \ref{Roman} for
the comparison of our results with those of Refs. \cite{INNa2002} and
\cite{RCCh2014}.

The L--T model obtained in this paper is the same as the one investigated in
Refs. \cite{YKNa2008} and \cite{Yoo2010}. Those authors took into account the
conditions imposed on the solutions of equations by the relations at the AH by a
method different in technical detail, but equivalent to the one employed here,
and calculated other functions for the resulting L--T model. Their radial
coordinate is different from ours, it is defined so that the equation of the
observer's past light cone is $t - t_{\rm obs} = - r$. Consequently, no
straightforward comparison of the results is possible. But they also found that
the parameters of the $\Lambda$CDM model uniquely define the energy and mass
functions in the associated L--T model with $t_B =$ constant.

The idea of drawing useful information from the values of various quantities at
the apparent horizon was first published by Hellaby \cite{Hell2006}. It was
applied here and in Ref. \cite{Kras2014} to several new examples. The results
might provide inspiration for investigations of less special models.

\appendix

\section{Derivation of (\ref{4.10})}\label{solve410}

\setcounter{equation}{0}

\subsection{$E > 0$}

In this case, $k < 0$ and ${\cal U}_0 > 1$.

The direct result of taking the limit $r \to 0$ in (\ref{3.16}) is, with use of
(\ref{3.17}) and (\ref{4.6}) -- (\ref{4.9}) and after simplifying
\begin{equation}\label{a.1}
F_1 F_2 = 0,
\end{equation}
where
\begin{eqnarray}
F_1 &\df& X - \sqrt{- k + \frac {2 M_0 H_0} X}, \label{a.2} \\
&& \nonumber \\
F_2 &\df& \sqrt{- k + \frac {2 M_0 H_0} X}\ \left(\frac {B_3} {B_1}\right)_0 -
X. \label{a.3}
\end{eqnarray}
The equation $F_1 = 0$ leads to (\ref{4.10}), so it has to be verified that
$F_2$ cannot be zero.

We substitute for $\left(B_3/B_1\right)_0$ from (\ref{4.5}) and rewrite
(\ref{4.4}) in the form
\begin{equation}\label{a.4}
- k = \frac {M_0 H_0} X\ \left({\cal U}_0 - 1\right).
\end{equation}
With (\ref{a.4}), the equation $F_2 = 0$ becomes
\begin{eqnarray}\label{a.5}
&&\hspace{-4mm} X \left\{\frac 3 2\ \frac {\sqrt{y + 1}} {(y - 1)^{3/2}}\
\left[\sqrt{y^2 - 1} - \ln \left(y + \sqrt{y^2 - 1}\right)\right] - 1\right\}
\nonumber \\
&& = 0,
\end{eqnarray}
where $y \df {\cal U}_0$. The solution $X = 0$ was excluded by assumption -- see
under (\ref{4.1}). The second factor in (\ref{a.5}) being zero is equivalent to
\begin{eqnarray}\label{a.6}
&&\hspace{-4mm} g(y) \df \sqrt{y^2 - 1} - \ln \left(y + \sqrt{y^2 - 1}\right) -
\frac 2 3\ \frac {(y - 1)^{3/2}} {\sqrt{y + 1}} \nonumber \\
&& = 0.
\end{eqnarray}
We have $g(1) = 0$ and
\begin{equation}\label{a.7}
\dr g y = \frac 1 3 \left(\frac {y - 1} {y + 1}\right)^{3/2},
\end{equation}
so, obviously, $\dril g y > 0$ for all $y > 1$, and, consequently, $g(y) > 0$
for $y > 1$. Thus, (\ref{a.6}) has no other solutions than $y = 1$. However,
note that $y > 1$ must hold, from (\ref{4.2}) (because $k < 0$ and $X > 0$).
Consequently, (\ref{4.10}) remains as the only acceptable consequence of
(\ref{a.1}). $\square$

\subsection{$E < 0$}

In this case, $0 < k < \infty$, $0 < X^3 < 2M_0 H_0$, $-1 < {\cal U}_0 < +1$,
and $(B_3/B_1)_0$ has to be replaced by $(\widetilde{B}_3/B_1)_0$ given by
(\ref{4.15}). So, using (\ref{a.4}) in the new $F_2$, we obtain instead of
(\ref{a.6})
\begin{equation}\label{a.8}
g_2(y) \df \arccos y - \sqrt{1 - y^2} - \frac 2 3\ \frac {(1 - y)^{3/2}}
{\sqrt{y + 1}} = 0.
\end{equation}
We have $g_2(-1) = - \infty$, $g_2(+1) = 0$, and
\begin{equation}\label{a.9}
\dr {g_2} y = \frac 1 3 \left(\frac {1 - y} {1 + y}\right)^{3/2},
\end{equation}
so $\dril {g_2} y > 0$ for all $y \in [-1, +1)$. Hence, (\ref{a.8}) has no other
solutions for $-1 \leq y \leq +1$ than $y = +1$. But $y = +1$ implies, via
(\ref{4.2}), $kX = 0$, which is impossible when $k > 0$ and $X > 0$. So, again,
(\ref{4.10}) is the only acceptable consequence of (\ref{a.1}). $\square$

\section{Derivation of (\ref{4.14})}\label{der418}

\setcounter{equation}{0}

We apply the de l'H\^{o}pital rule to the last term in (\ref{4.13}), then use
(\ref{2.21}), (\ref{3.8}), (\ref{2.29}), (\ref{2.27}), (\ref{4.1}), (\ref{4.6})
and (\ref{4.10}). In the resulting expression, several terms can be readily
calculated. Only one nontrivial limit remains:
\begin{widetext}
\begin{eqnarray}\label{b.1}
&& \lim_{r \to 0} \left[\frac 1 r \left(\sqrt{1 + 2E} + A_1 B_1 / {\cal
D}\right)\right] = \tfrac 3 2 \Omega_m X - \frac {M_0 H_0} {X^2} - \frac 1 {2
X^2} \lim_{r \to 0} \dr {\cal F} r \nonumber \\
&&\ \ \ \ \  + \left(\frac 1 X + \frac {M_0 H_0} {X^4}\right) \lim_{r \to 0}
\left\{\frac 1 r \left[\frac {\dril z r} {\sqrt{\Omega_m (1 + z)^3 +
\Omega_{\Lambda}}} - \frac {\cal D} r\right]\right\}.
\end{eqnarray}
Now we substitute (\ref{b.1}) in (\ref{4.13}) and solve the result for $\lim_{r
\to 0} \dril {\cal F} r$:
\begin{eqnarray}\label{b.2}
&& \left[\frac 1 k\ \left(X + \frac {M_0 H_0} {X^2}\right) \left(\frac {B_3}
{B_1}\right)_0 - \frac {3X} {2k} + \frac 1 {2X}\right] \lim_{r \to 0} \dr {\cal
F} r \nonumber \\
&& = \left(\tfrac 3 2 \Omega_m - 1\right) X^2 - \frac {M_0 H_0} X + \left(1 +
\frac {M_0 H_0} {X^3}\right) \lim_{r \to 0} \left\{\frac 1 r \left[\frac {\dril
z r} {\sqrt{\Omega_m (1 + z)^3 + \Omega_{\Lambda}}} - \frac {\cal D}
r\right]\right\}.
\end{eqnarray}
In the last term above we substitute for $\dril z r$ from (\ref{3.16}), then for
$E$ from (\ref{2.21}). Several terms can again be readily calculated. In the
remaining limit we use (\ref{2.29}) to eliminate the large square root. The
result is
\begin{eqnarray}\label{b.3}
&& \lim_{r \to 0} \left\{\frac 1 r \left[\frac {\dril z r} {\sqrt{\Omega_m (1 +
z)^3 + \Omega_{\Lambda}}} - \frac {\cal D} r\right]\right\} \nonumber  \\
&& = X^2 + \frac 1 k\ \left[- \frac X 2 + \frac {M_0 H_0} {X^2} \left(\frac
{B_3} {B_1}\right)_0\right] \lim_{r \to 0} \dr {\cal F} r - \lim_{r \to 0}
\left[\frac 1 r \left(\frac {A_1 B_1} {r \sqrt{1 + 2E}} + \frac {\cal D}
r\right)\right].
\end{eqnarray}
Here, using (\ref{3.8}) for $B_1$ and (\ref{2.21}) for $E$, we again apply the
de l'H\^{o}pital rule to calculate
\begin{eqnarray}\label{b.4}
&& \lim_{r \to 0} \left[\frac 1 r \left(\frac {A_1 B_1} {r \sqrt{1 + 2E}} +
\frac {\cal D} r\right)\right] \nonumber  \\
&& =  \tfrac 3 2 \Omega_m X^2 - \frac {M_0 H_0} X - \frac 1 {2 X} \lim_{r \to
0} \dr {\cal F} r + \left(1 + \frac {M_0 H_0} {X^3}\right) \lim_{r \to 0}
\left\{\frac 1 r \left[\frac {\dril z r} {\sqrt{\Omega_m (1 + z)^3 +
\Omega_{\Lambda}}} - \frac {\cal D} r\right]\right\}.
\end{eqnarray}
Substituting  (\ref{b.4}) in (\ref{b.3}) we get
\begin{eqnarray}\label{b.5}
&& \left[\frac {M_0 H_0} {k X^2} \left(\frac {B_3} {B_1}\right)_0 - \frac {X}
{2k} + \frac 1 {2X}\right] \lim_{r \to 0} \dr {\cal F} r \nonumber \\
&&= \left(\tfrac 3 2 \Omega_m - 1\right) X^2 - \frac {M_0 H_0} X + \left(2 +
\frac {M_0 H_0} {X^3}\right) \lim_{r \to 0} \left\{\frac 1 r \left[\frac {\dril
z r} {\sqrt{\Omega_m (1 + z)^3 + \Omega_{\Lambda}}} - \frac {\cal D}
r\right]\right\}.
\end{eqnarray}
\end{widetext}
Equations (\ref{b.2}) and (\ref{b.5}) determine $\lim_{r \to 0} \dril {\cal F}
r$ as in (\ref{4.14}), using (\ref{4.10}) and (\ref{4.11}). $\square$

\section{Proof that $\dril T X > 0$ for $X > 0$ in Sec.
\ref{deterXandk}}\label{limproofs}

\setcounter{equation}{0}

We substitute in (\ref{6.2}) for ${\cal U}_0$ from (\ref{4.2}) and
\begin{equation}\label{c.1}
k = 2 M_0 H_0 / X - X^2
\end{equation}
from (\ref{4.11}), and calculate
\begin{equation}\label{c.2}
\dr {T_-} X = \frac {3 M_0 \left(X + M_0 H_0 / X^2\right)} {\left(X^2 - 2 M_0
H_0 / X\right)^{5/2}}\ F(X),
\end{equation}
where
\begin{equation}\label{c.3}
F(X) \df \ln \left({\cal U}_0 + \sqrt{{{\cal U}_0}^2 - 1}\right) - 3\ \frac
{\sqrt{X^2 - 2 M_0 H_0 / X}} {X + M_0 H_0 / X^2},
\end{equation}
with
\begin{equation}\label{c.4}
{\cal U}_0 = \frac {X^3} {M_0 H_0} - 1.
\end{equation}
We have
\begin{eqnarray}
&& F((2M_0 H_0)^{1/3}) = 0, \label{c.5} \\
&& \lim_{X^3 \to 2M_0 H_0} \dr {T_-} X = \frac {2^{2/3} M_0} {5 \left(M_0
H_0\right)^{4/3}}, \label{c.6} \\
&& \dr F X = \frac {3 \left(X^2 - 2 M_0 H_0 / X\right)^{3/2}} {X^2 \left(X + M_0
H_0 / X^2\right)^2}, \label{c.7}
\end{eqnarray}
so $\dril F X > 0$ for all $X > (2M_0 H_0)^{1/3}$. Equations (\ref{c.7}) and
(\ref{c.5}) show that $F > 0$ for all $X > (2M_0 H_0)^{1/3}$, and then
(\ref{c.2}) shows that $\dril {T_-} X > 0$ in the same range.

Doing analogous operations in (\ref{6.4}) we obtain
\begin{eqnarray}
&& \dr {T_+} X = \frac {3 M_0 \left(X + M_0 H_0 / X^2\right)} {\left(2 M_0 H_0
/ X - X^2\right)^{5/2}}\ G(X), \label{c.8} \\
&& G(X) \df \arccos {\cal U}_0 - 3\ \frac {\sqrt{2 M_0 H_0 / X - X^2}} {X + M_0
H_0 / X^2}, \label{c.9} \\
&& \lim_{X^3 \to 2M_0 H_0} \dr {T_+} X = \frac {2^{2/3} M_0} {5 \left(M_0
H_0\right)^{4/3}}, \label{c.10} \\
&& \dr G X = - \frac {3 \left(2 M_0 H_0 / X - X^2\right)^{3/2}} {X^2 \left(X +
M_0 H_0 / X^2\right)^2}, \label{c.11}
\end{eqnarray}
so $\dril G X < 0$ for $0 < X < (2M_0 H_0)^{1/3}$. We also have
\begin{eqnarray}
&& G(0) = \arccos (-1) = \pi, \label{c.12} \\
&& G((2M_0 H_0)^{1/3}) = 0. \label{c.13}
\end{eqnarray}
This means that for $0 < X < (2M_0 H_0)^{1/3}$ the function $G(X)$ uniformly
decreases from $\pi$ to zero, so $G(X) > 0$ in this whole interval.
Consequently, in (\ref{c.8}), $\dril {T_+} X > 0$ in this interval. Then, from
(\ref{6.7}) and $T_+(0) = 0$, it follows that in this interval $T_+$ is
everywhere smaller than the $T_0$ from (\ref{6.5}). $\square$

\section{Comparison of (\ref{6.2}) -- (\ref{6.4}) to the result of Iguchi et al.
\cite{INNa2002}}\label{Iguchi}

\setcounter{equation}{0}

Iguchi et al. used different units and did not refer directly to the age of the
model universe $T_-$ or $T_+$. Instead, they referred to $\Omega_0$ -- the ratio
of the central density to the RW critical density, which determines the age of
the model via an equation that can be solved only numerically. So, the
comparison cannot be done by directly comparing numbers.

Our numerical time unit (\ref{2.36}) followed from assuming $H_0 = 6.71$ in
(\ref{2.35}). They assumed $H_0 = 1$, so their numerical time unit is
\begin{eqnarray}\label{d.1}
&& 1\ {\rm NTU}_{Iguchi} = c / 67.1 = 0.447094 \times 10^4\ {\rm Mpc} \nonumber
\\
&& = 0.149\ {\rm NTU}_{Kras} \equiv (1/6.71)\ {\rm NTU}_{Kras}.
\end{eqnarray}

They calculated numerically the functions $E(z)$ for different values of
$\Omega_0$. (Our $k$ is their $- 2E(0)$, see their (2.1) vs our (\ref{2.1}) and
(\ref{2.21}).) Thus, in effect, they treated the age of the model universe as a
free parameter and did the numerical calculations for different assumed values
of $T_-$. The highest value used in their paper, $\Omega_0 = 1.0$, means that
the central density is equal to critical. Consequently, in this case, their
$E(0) = 0$, so our $k = 0$. From our (\ref{4.11}) it follows that then $X =
(2M_0 H_0)^{1/3}$, and our (\ref{6.7}) implies that the age of the model
universe is $T_- = 0.0993541977$ NTU. Calculating the corresponding $k$ from
(\ref{6.2}) using (\ref{4.2}) and (\ref{4.11}) we obtain $k = -1.392464 \times
10^{-3}$, which is as close to zero as the numerical precision allows (note,
from (\ref{6.2}), that calculating $k$ given $T_-$ in a neighbourhood of $k = 0$
requires evaluating an expression of the form 0/0).

The smallest $\Omega_0$ used in Ref. \cite{INNa2002} is 0.1. Figure 4 in Ref.
\cite{INNa2002} indicates that then their $E(0) \approx 0.42$, which corresponds
to our $k \approx - 0.84$. Taking this value we find $X = 2.4941229$ from
(\ref{4.11}), and then, from (\ref{6.2}), $T_- = 0.1022$ NTU $= 10.0156 \times
10^9$ y.

However, as stated in the paragraph below our (\ref{6.4}), the condition
$z(r_{\rm AH}) = z_{\rm AH}$ uniquely fixes $k$; the only uncertainty about
the value of $k$ may come from numerical problems. With $k$ given, the age of
the model universe, (\ref{6.2}) or (\ref{6.4}), is also fixed. Consequently, it
is not correct to treat this age as a free parameter -- there is just one L--T
model to be compared with $\Lambda$CDM.

\section{Comparing the $\Lambda$CDM and L--T models}\label{correctr}

\setcounter{equation}{0}

Equation (\ref{2.26}) applies also in the $\Lambda$CDM model (where, in fact, it
is an identity), with $R(t, r) = r S(t)$; the $S(t)$ is the $\Lambda$CDM scale
factor. The same is true for (\ref{2.25}) at the AH. Recall that the values of
$z_{\rm AH}$ and ${\cal D}_{\rm AH}$, given by (\ref{2.32}) and (\ref{2.33}),
are determined by the right-hand side of (\ref{2.26}), and are independent of
the algebraic form of $R$. Hence, they will be the same in the $\Lambda$CDM and
L--T models. Therefore, (\ref{2.30}) also applies in the $\Lambda$CDM limit.
Consequently, if $M_0$ is chosen the same in the $\Lambda$CDM and L--T models,
the $r_{\rm AH}$ will also have the same value in both models. The conclusion is
that if the $\Lambda$CDM metric is represented in the form (\ref{2.12}), then,
by applying a linear transformation to $r$, one can assure that at the AH $r$ is
the same in both models and $z$ is the same in both models.

The function $z(r)$ in the $\Lambda$CDM model is calculated as follows:

1. Solve the null geodesic equation for (\ref{2.12}) to find (numerically)
$t(r)$ along the geodesic.

2. Use (\ref{2.13}) for $z(t)$, where $t_o = 0$ is the observation time and
$t_e$ is the running value of $t$.

3. Use the $z(t)$ function and the $t(r)$ table to find the $z(r)$ table.

This $z(r)$ is {\em not} guaranteed to obey $z(r_{\rm AH}) = z_{\rm AH}$, where
$r_{\rm AH}$ and $z_{\rm AH}$ are taken from the L--T model. This is the point
that was not taken care of in Ref. \cite{Kras2014}. It was assumed there that
the two $r$-coordinates are the same, but they were not. However, all the
qualitative conclusions from the comparison of the two light cones formulated
there remain correct.

In order to make the two $r$-coordinates compatible, one must apply the
transformation $r = Cr'$ to the $r$ of $\Lambda$CDM and choose the constant $C$
so that the $z(r)$ curves of the two models both pass through the point $(r, z)
= (r_{\rm AH}, z_{\rm AH})$. This is how both panels in our Fig.
\ref{drawzfullrange} were constructed. The $r$-coordinate of the $\Lambda$CDM
model was transformed in the same way in Fig. \ref{conefirst}.

\section{Comparison of the results of Romano et al. \cite{RCCh2014} to
ours}\label{Roman}

\setcounter{equation}{0}

Similar to Ref. \cite{INNa2002}, the authors of Ref. \cite{RCCh2014} treated $k$
as a free parameter to be adjusted to observations. The relations between their
parameters and ours are the following. Their $H_0$ coincides with our $H_0$,
except for the units. Their $r$, $E$ and $R$ coincide with ours. From their (7),
(21) and (23) it follows that their
\begin{equation}\label{f.1}
a_0 = \lim_{r \to 0} (R/r).
\end{equation}
{}From our (\ref{2.26}), (\ref{2.27}) and (\ref{4.1}) it follows that

\begin{equation}\label{f.2}
{\rm their}\ a_0 = {\rm our}\ X/H_0.
\end{equation}
Then, from their (7), (21), (27) and (32) it follows that
\begin{equation}\label{f.3}
{\rm their}\ (k_0, K_0) = {\rm our}\ (k, k/X).
\end{equation}
So, our $k$ and $X$ given by (\ref{9.1}) and (\ref{9.2}) translate to their $K_0
\approx -4.4169$. But this value is not in the set $K_0 \in \{-0.9376, -0.91\}$,
for which graphs were drawn in Ref. \cite{RCCh2014}. Hence, there is no common
subset of our results and theirs.


\begin{thebibliography}{99}
\bibitem{Lema1933} G. Lema\^{\i}tre, {\it Ann. Soc. Sci. Bruxelles} {\bf A53},
    51 (1933); English translation, with historical comments: {\it Gen.
    Relativ. Gravit.} {\bf 29}, 637 (1997).

\bibitem{Tolm1934} R. C. Tolman, {\it Proc. Nat. Acad. Sci. USA} {\bf 20}, 169
    (1934); reprinted, with historical comments: {\it Gen. Relativ. Gravit.}
    {\bf 29}, 931 (1997).

\bibitem{Kras2014} A. Krasi\'nski, {\it Phys. Rev.} {\bf D89}, 023520 (2014).

\bibitem{INNa2002} H. Iguchi, T. Nakamura and K. Nakao, {\it Progr. Theor.
    Phys}. {\bf 108}, 809 (2002).

\bibitem{PlKr2006} J. Pleba\'nski and A. Krasi\'nski, {\it An Introduction to
    General Relativity and Cosmology}. Cambridge University Press 2006, 534 pp,
    ISBN 0-521-85623-X.

\bibitem{Kras1997} A. Krasi\'nski, {\it Inhomogeneous Cosmological Models},
    Cambridge University Press 1997, 317 pp, ISBN 0 521 48180 5.

\bibitem {Bond1947} H. Bondi, {\it Mon. Not. Roy. Astr. Soc.} {\bf 107}, 410
    (1947); reprinted as a Golden Oldie in {\it Gen. Relativ. Gravit.} {\bf
31}, 1777 (1999).

\bibitem{BKHC2010} K. Bolejko, A. Krasi\'nski, C. Hellaby and M.-N.
    C\'el\'erier, {\it Structures in the Universe by exact methods: formation,
    evolution, interactions}. Cambridge University Press 2010, 242 pp, ISBN
    978-0-521-76914-3.

\bibitem{Plan2013} Planck collaboration, {\it Planck 2013 results. XVI.
    Cosmological parameters}. arXiv 1303.5076; accepted for {\it
    Astronomy and Astrophysics}

\bibitem{HeLa1985} C. Hellaby and K. Lake, {\it Astrophys. J.} {\bf 290}, 381
    (1985) + erratum {\it Astrophys. J.} {\bf 300}, 461 (1986).

\bibitem{KrHe2004} A. Krasi\'nski and C. Hellaby, {\it Phys. Rev.} {\bf D69},
    043502 (2004).

\bibitem{Elli1971} G. F. R. Ellis, in {\it Proceedings of the International
    School of Physics `Enrico Fermi', Course 47: General Relativity and
    Cosmology}, ed. R. K. Sachs. Academic Press, New York and London (1971), pp.
    104 -- 182; reprinted as a Golden Oldie in {\it Gen. Relativ. Gravit.} {\bf
    41}, 581 (2009).

\bibitem{Szek1980} P. Szekeres, in: {\it     Gravitational Radiation, Collapsed
    Objects and Exact Solutions}. Edited by C. Edwards. Springer (Lecture Notes
    in Physics, vol. 124), New York, pp. 477 -- 487 (1980).

\bibitem{HeLa1984} C. Hellaby and K. Lake, {\it Astrophys. J.} {\bf 282}, 1
    (1984) + erratum {\it Astrophys. J.} {\bf 294}, 702 (1985).

\bibitem{unitconver} http://www.asknumbers.com/LengthConversion.aspx

\bibitem{CBKr2010} M.-N. C\'el\'erier, K. Bolejko and A. Krasi\'nski, {\it
    Astronomy and Astrophysics} {\bf 518}, A21 (2010).

\bibitem{Luci2004} M. Luciuk, Astronomical Redshift,\\
http://www.asterism.org/tutorials/tut29-1.htm, last updated 2004.

\bibitem{Ries1998} A. G. Riess {\it et al.}, {\it Astron. J}. {\bf 116}, 1009
    (1998).

\bibitem{Perl1999} S. Perlmutter {\it et al.}, {\it Astrophys. J.} {\bf 517},
    565 (1999).

\bibitem{Jone2014} D. O. Jones {\it et al.}, {\it Astrophys. J.} {\bf 768}, 166 (2013).

\bibitem{RCCh2014} A. E. Romano, H.-W. Chiang and P. Chen, {\it Class. Quant.
    Grav.} {\bf 31}, 115008 (2014).

\bibitem{YKNa2008} C.-M. Yoo, T. Kai and K. Nakao, {\it Progr. Theor. Phys.}
    {\bf 120}, 937 (2008).

\bibitem{Yoo2010} C.-M. Yoo, {\it Progr. Theor. Phys.} {\bf 124}, 645 (2010).

\bibitem{Hell2006} C. Hellaby, {\it Mon. Not. Roy. Astron. Soc.} {\bf 370}, 239
(2006).
\end{thebibliography}
\end{document}